\documentclass[letterpaper,11pt]{article}

\usepackage{microtype}

\usepackage{epic,eepic,amsmath,latexsym,amssymb,color}
\usepackage{ifthen,graphics,epsfig}
\usepackage{times, comment}

\usepackage{array}
\usepackage{color}
\usepackage{colortbl}
\usepackage{multirow}
\usepackage{pgf}
\usepackage{tikz}
\usetikzlibrary{arrows,matrix, calc, fit}
\usepackage{ifthen}

\usepackage{rotating}

\usepackage{pgfplots}

\usepackage{comment}
\usepackage{caption}

\usepackage{intcalc}
\usetikzlibrary{patterns}
\usepackage{floatrow}
\title{Optimal Self-Stabilizing Mobile Byzantine-Tolerant Regular Register with bounded timestamps}

\author{Silvia Bonomi$^\star$, Antonella Del Pozzo$^\star$,\\ Maria Potop-Butucaru$^\dagger$, S\'ebastien Tixeuil$^\dagger$\\~\\
	$^\star$Sapienza Universit\`{a} di Roma,Via Ariosto 25, 00185 Roma, Italy\\
	\texttt{\{bonomi, delpozzo\}}$@$dis.uniroma1.it\\
	$^\dagger$Sorbonne Universit\'e, CNRS, Laboratoire d'Informatique de Paris 6,\\ F-75005 Paris, France\\
	\texttt{\{maria.potop-butucaru, sebastien.tixeuil\}}$@$lip6.fr}

\begin{document}

	 \def\startCirc#1{\tikz[remember picture,overlay]\path node[inner sep=0, anchor=south] (st) {#1} coordinate (start) at (st.center);}%
	 \def\endCirc#1{\tikz[remember picture,overlay]\path node[inner sep=0, anchor=south] (en) {#1} coordinate (end) at (en.center);%
	 	\begin{tikzpicture}[overlay, remember picture]%
	 	\path (start);%
	 	\pgfgetlastxy{\startx}{\starty}%
	 	\path (end);%
	 	\pgfgetlastxy{\endx}{\endy}%
	 	\pgfmathsetlengthmacro{\xdiff}{\endx-\startx}%
	 	\pgfmathsetlengthmacro{\ydiff}{\endy-\starty}%
	 	\pgfmathtruncatemacro{\xdifft}{\xdiff}%
	 	\pgfmathsetmacro{\xdiffFixed}{ifthenelse(equal(\xdifft,0),1,\xdiff)}%
	 	\pgfmathsetmacro{\angle}{ifthenelse(equal(\xdiffFixed,1),90,atan(\ydiff/\xdiffFixed))}%
	 	\pgfmathsetlengthmacro{\xydiff}{sqrt(abs(\xdiff^2) + abs(\ydiff^2))}%
	 	\path node[draw,rectangle,red, rounded corners=2mm, rotate=\angle, minimum width=\xydiff+4ex, minimum height=2.5ex] at ($(start)!.5!(end)$) {};%
	 	\end{tikzpicture}%
	 }
	 
	 \newlength {\squarewidth}
	 \renewenvironment {square}
	 {
	 	\setlength {\squarewidth} {\linewidth}
	 	\addtolength {\squarewidth} {-12pt}
	 	\renewcommand{\baselinestretch}{0.75} \footnotesize
	 	\begin {center}
	 	\begin {tabular} {|c|} \hline
	 	\begin {minipage} {\squarewidth}
	 	\medskip
	 }{
	 \end {minipage}
	 \\ \hline
	\end{tabular}
\end{center}
}



  \def\startCirc#1{\tikz[remember picture,overlay]\path node[inner sep=0, anchor=south] (st) {#1} coordinate (start) at (st.center);}%
  \def\endCirc#1{\tikz[remember picture,overlay]\path node[inner sep=0, anchor=south] (en) {#1} coordinate (end) at (en.center);%
    \begin{tikzpicture}[overlay, remember picture]%
      \path (start);%
      \pgfgetlastxy{\startx}{\starty}%
      \path (end);%
      \pgfgetlastxy{\endx}{\endy}%
      \pgfmathsetlengthmacro{\xdiff}{\endx-\startx}%
      \pgfmathsetlengthmacro{\ydiff}{\endy-\starty}%
      \pgfmathtruncatemacro{\xdifft}{\xdiff}%
      \pgfmathsetmacro{\xdiffFixed}{ifthenelse(equal(\xdifft,0),1,\xdiff)}%
      \pgfmathsetmacro{\angle}{ifthenelse(equal(\xdiffFixed,1),90,atan(\ydiff/\xdiffFixed))}%
      \pgfmathsetlengthmacro{\xydiff}{sqrt(abs(\xdiff^2) + abs(\ydiff^2))}%
      \path node[draw,rectangle,red, rounded corners=2mm, rotate=\angle, minimum width=\xydiff+4ex, minimum height=2.5ex] at ($(start)!.5!(end)$) {};%
    \end{tikzpicture}%
  }




\newtheorem{definition}{Definition}
\newtheorem{theorem}{Theorem}
\newtheorem{lemma}{Lemma}
\newtheorem{corollary}{Corollary}
\newtheorem{property}{Property}
\newcommand{\note}[1]{\noindent\textcolor{red}{{\fontfamily{phv}\selectfont NOTE: #1}}}
\newcommand{\toto}{xxx}
\newenvironment{proofT}{\noindent{\bf
Proof }} {\hspace*{\fill}$\Box_{Theorem~\ref{\toto}}$\par\vspace{3mm}}
\newenvironment{proofL}{\noindent{\bf
Proof }} {\hspace*{\fill}$\Box_{Lemma~\ref{\toto}}$\par\vspace{3mm}}
\newenvironment{proofC}{\noindent{\bf
Proof }} {\hspace*{\fill}$\Box_{Corollary~\ref{\toto}}$\par\vspace{3mm}}

\newcounter{linecounter}
\newcommand{\linenumbering}{(\arabic{linecounter})}
\renewcommand{\line}[1]{\refstepcounter{linecounter}
\label{#1}
\linenumbering}
\newcommand{\resetline}{\setcounter{linecounter}{0}}


\newcommand{\vir}[1]{``#1''}

\newcommand{\ceil}[1]{\lceil #1 \rceil}
\newcommand{\op}[1]{${\sf #1}()$}
\newcommand{\opArg}[2]{{\sf #1}(#2)}
\newcommand{\msg}[1]{{\sc #1}$()$}
\newcommand{\msgArg}[2]{{\sc #1}$(#2)$}
\newcommand{\ang}[2]{\langle #1, #2 \rangle}
\newcommand{\tb}[1]{$t_B(op_{#1})$}
\newcommand{\te}[1]{$t_E(op_{#1})$}
\newcommand{\Mod}[1]{\ (\text{mod}\ #1)}


%

\newcommand{\intervallo}[3]{ 
	\draw[-] (#1,-.5) -- (#1+\deltaPiccolo,-.5);	
	\draw [-] (#1+\deltaPiccolo,-.5) -- (#1+#2*\deltaPiccolo,-.5);	
	\foreach \x in {0,...,#2}
	\draw[] (#1+\x*\deltaPiccolo, \n) -- (#1+\x*\deltaPiccolo, -.6);
	\node[]() at (#1, -.75) {\scriptsize{$#3$}};
	\node[]() at (#1+#2*\deltaPiccolo+.1, -.75) {\scriptsize{$#3+#2\delta$}};
}

\newcommand{\lettura}[1]{
	\draw[|-|] (#1,-.5) -- (#1+\deltaPiccolo,-.5);	
	\draw [-|] (#1+\deltaPiccolo,-.5) -- (#1+2*\deltaPiccolo,-.5);
	\node[]() at (#1+\deltaPiccolo, -.7) {\scriptsize{\op{read}}};
	
	\foreach \x in {0,...,2}
	\draw[] (#1+\x*\deltaPiccolo, -.7) -- (#1+\x*\deltaPiccolo, \n);
}

\newcommand{\letturaLunga}[2]{ 
	\draw[|-|] (#1,-.5) -- (#1+\deltaPiccolo, -.5);	
	\draw [-|] (#1+\deltaPiccolo, -.5) -- (#1+#2*\deltaPiccolo, -.5);
	
	\foreach \x in {0,...,#2}
	\draw[] (#1+\x*\deltaPiccolo, \n) -- (#1+\x*\deltaPiccolo, -.7);
}

\newcommand{\scrittura}[1]{
	\draw[|-|] (#1,-1) -- (#1+\deltaPiccolo,-1);	
	\node[] () at (#1+\deltaPiccolo/2, -1.2) {\op{write}};
	\foreach \x in {0,...,1}
	\draw[dotted] (#1+\x*\deltaPiccolo, -1) -- (#1+\x*\deltaPiccolo, \n+.5);
}

\newcommand{\scritturaValore}[2]{
	\draw[|-|] (#1,-1) -- (#1+\deltaPiccolo,-1);	
	\node[] () at (#1+\deltaPiccolo/2, -1.2) {\scriptsize ${\sf w}(#2)$};
	\foreach \x in {0,...,1}
	\draw[dotted] (#1+\x*\deltaPiccolo, -1) -- (#1+\x*\deltaPiccolo, \n+.5);
}

\newcommand{\scritturaValoreAltezza}[3]{
	\draw[|-|, thick] (#1,#3) -- (#1+\deltaPiccolo,#3);	
	\node[] () at (#1+\deltaPiccolo/2, #3-.2) {\scriptsize ${\sf w}(#2)$};
	\foreach \x in {0,...,1}
	\draw[dotted] (#1+\x*\deltaPiccolo, #3) -- (#1+\x*\deltaPiccolo, \n+.5);
}

\newcommand{\lineaVerticale}[4]{
	\draw[#4] (#1, #2) -- (#1, \n+.5);
	\node[] () at (#1, #2-.2) {#3};
}

\newcommand{\lineaVerticaleLabelBasso}[5]{
	\draw[#4] (#1, #2) -- (#1, #5);
	\node[] () at (#1, #5+.2) {#3};
}

\newcommand{\reply}[3]{
	\draw[->, #3, thick](#1,#2)--(#1+.2,#2-.25);
}

\newcommand{\bc}[3]{
	\draw[<->, #3, thick](#1,#2+.25)--(#1,#2-.25);
}

\newcommand{\req}[3]{
	\draw[->, #3, thick] (#1,#2-.25) -- (#1+.2,#2);
}

\newcommand{\punto}[4]{
	\fill[#3] (#1,#2) circle[radius=2pt];
	\node[] () at (#1,#2-.2) {#4};
}

\newcommand{\curato}[2]{
	\filldraw[fill=gray!40!white, draw=black] (#1,#2) rectangle (#1+\gammaCuring,#2-.2);
}

\newcommand{\rettangoloOpaco}[4]{
	\filldraw[fill=#4!40!white, fill opacity=0.1, draw=black] (#1,#2) rectangle (#1+#3,#2-.4);
}

\newcommand{\curatoParziale}[4]{
	\filldraw[fill=#4!40!white, draw=black] (#1,#2) rectangle (#1+#3,#2-.2);
}

\newcommand{\processes}{ 
	\foreach \x in {0,...,\n} 
	\draw[->] (0,\x) -- (\lenght,\x); 
	\foreach \x in {0,..., \n}
	\node[]() at (-.3,\x) {$s_{\x}$};	
}

\newcommand{\processesDaUno}{ 
	\foreach \x in {1,...,\n} 
	\draw[->] (0,\x) -- (\lenght,\x); 
	\foreach \x in {1,..., \n}
	\node[]() at (-.3,\x) {$s_{\x}$};	
}

\newcommand{\processesSilvia}{ 
	\foreach \x in {0,...,\n} 
	\draw[->] (0,\x) -- (\lenght,\x); 
	\foreach \x in {0,..., \n}
	\node[]() at (-.3,\x) {$s_{\intcalcMod{\n-\x}{\n+1}}$}; 	
}

\newcommand{\values}[1]{
	\node[]() at (2, \n+2.7) {Model: #1, $\delta= \deltaPiccolo$, $\Delta= \deltaGrande$, $k=$ \pgfmathparse{int((\deltaGrande) / (\deltaPiccolo*2))}\pgfmathresult , $\gamma= \gammaCuring$};
}

\newcommand{\collected}[2]{
	\node[]() at (1.5, \n+2) {Correct replies: #1; Incorrect replies: {\color{red}}#2};	
}	
\newcommand{\faults}[5]{ 
	\foreach \x in {0,...,#3}
	\filldraw[fill=red!40!white, draw=black] (#1+\x*\deltaGrande,\intcalcMod{ #2+\x}{\n+1}) rectangle (#1+\deltaGrande+\x*\deltaGrande,\intcalcMod{ #2+\x}{\n+1}-.3);
	
	\foreach \x in {0,...,\intcalcSub{#3}{1}}
	\filldraw[fill=#4!40!white, draw=black] (#1+\deltaGrande+\x*\deltaGrande,\intcalcMod{#2+\x}{\n+1}) rectangle (#1+\deltaGrande+\x*\deltaGrande+\gammaCuring,\intcalcMod{#2+\x}{\n+1}-.2);
	
	\ifnum#5>0 
	\foreach \x in {0,...,#5}
		\filldraw[fill=#4!40!white, draw=black] (0,\intcalcMod{#2-\x+\n}{\n+2}) rectangle (\gammaCuring-\x*\deltaGrande,\intcalcMod{#2-\x+\n}{\n+2}-.2);	
	\fi
}

\newcommand{\faultsInverso}[5]{ 
	\foreach \x in {0,...,#3}
	\filldraw[fill=red!40!white, draw=black] (#1+\x*\deltaGrande,\intcalcMod{ #2-\x}{\n}) rectangle (#1+\deltaGrande+\x*\deltaGrande,\intcalcMod{ #2-\x}{\n}-.3);
	
	\foreach \x in {0,...,\intcalcSub{#3}{1}}
	\filldraw[fill=#4!40!white, draw=black] (#1+\deltaGrande+\x*\deltaGrande,\intcalcMod{#2-\x}{\n}) rectangle (#1+\deltaGrande+\x*\deltaGrande+\gammaCuring,\intcalcMod{#2-\x}{\n}-.2);
	
	\ifnum#5>0 
	\foreach \x in {0,...,#5}
	\filldraw[fill=#4!40!white, draw=black] (0,\intcalcMod{#2-\x+\n}{\n+2}) rectangle (\gammaCuring-\x*\deltaGrande,\intcalcMod{#2-\x+\n}{\n+2}-.2);	
	\fi
}

\newcommand{\faultsAdaptive}[5]{ 
	\foreach \x in {0,...,#3}
	
		\ifthenelse{{#1+\deltaGrande+\x*\deltaGrande = #5}}
			{\filldraw[fill=#4!40!white, draw=black] (#1+\x*\deltaGrande,\intcalcMod{#2+\x}{\n+1}) rectangle (#1+\deltaGrande+\x*\deltaGrande,\intcalcMod{#2+\x}{\n+1}+.3);}
			{\ifthenelse{{#1+\deltaGrande+\x*\deltaGrande < #5}}
				{\filldraw[fill=#4!40!white, draw=black] (#1+\x*\deltaGrande,\intcalcMod{#2+\x}{\n+1}) rectangle (#1+\deltaGrande+\x*\deltaGrande,\intcalcMod{#2+\x}{\n+1}+.3);}
				{\filldraw[fill=#4!40!white, draw=black] (#1+\x*\deltaGrande,\intcalcMod{#2+\x}{\n+1}) rectangle (#5,\intcalcMod{#2+\x}{\n+1}+.3);}}
	\foreach \x in {0,...,#3}
	\filldraw[fill=gray!40!white, draw=black] (#1+\deltaGrande+\x*\deltaGrande,\intcalcMod{#2+\x}{\n+1}) rectangle (#1+\deltaGrande+\x*\deltaGrande+\gammaCuring,\intcalcMod{#2+\x}{\n+1}+.2);
}

\newcommand{\faultsDelta}[5]{
	\foreach \x in {0,...,#3}
	\draw[pattern=#4] (#1+\x*#5,\intcalcMod{#2+\x}{\n+1}) rectangle (#1+#5+\x*#5,\intcalcMod{#2+\x}{\n+1}+.3);
	\foreach \x in {0,...,#3}
	\filldraw[fill=gray!40!white, draw=black] (#1+#5+\x*#5,\intcalcMod{#2+\x}{\n+1}) rectangle (#1+#5+\x*#5+\gammaCuring,\intcalcMod{#2+\x}{\n+1}+.2);
}


\maketitle

\begin{abstract}
This paper proposes the first implementation of a self-stabilizing regular register emulated by $n$ servers that is tolerant to both \emph{mobile Byzantine agents}, and \emph{transient failures}  in a round-free synchronous model. 
Differently from existing Mobile Byzantine tolerant register implementations, this paper considers a more powerful adversary where (i) the message delay (i.e., $\delta$) and the period of mobile Byzantine agents movement (i.e., $\Delta$) are completely decoupled and (ii) servers are not aware of their state i.e., they do not know if they have been corrupted or not by a mobile Byzantine agent.

The proposed protocol tolerates \emph{(i)} any number of transient failures, and \emph{(ii)} up to $f$ Mobile Byzantine agents. In addition, our implementation uses bounded timestamps from the $\mathcal{Z}_{13}$ domain and it is optimal with respect to the number of servers needed to tolerate $f$ mobile Byzantine agents in the given model.
~\\

%
%
%

\end{abstract}

%
\section{Introduction}
Byzantine fault tolerance is a fundamental building block in distributed system as Byzantine failures include all possible faults, attacks, virus infections  and  arbitrary behaviors that can occur in practice (even unforeseen ones). 
Such bad behaviors have been typically abstracted by assuming an upper bound $f$ on the number of Byzantine failures in the system. However, such assumption has two main limitations: (i) it is not suited for long lasting executions and (ii) it does not consider the fact that compromised processes/servers may be restored as infections may be blocked and confined or rejuvenation mechanisms can be put in place \cite{SBAM09} making the set of faulty processes changing along time.


Mobile Byzantine Failure (MBF) models have been recently introduced to integrate those concerns. Failures are represented by Byzantine agents that are managed by an omniscient adversary that ``moves'' them from a host process to another and when an agent is in some process it is able to corrupt it in an unforeseen manner. Models investigated so far in the context of mobile Byzantine failures consider mostly \emph{round-based} computations, and can be classified according to Byzantine mobility constraints: \emph{(i)} constrained mobility~\cite{Garay+95+ORA} agents may only move from one host to another when protocol messages are sent (similarly to how viruses would propagate), while \emph{(ii)} unconstrained mobility~\cite{Banu+2012,BDNP14,Garay+1994,Ostrovsky+91,Reischuk+85,Sasaki+2013} agents may move independently of protocol messages. In the case of unconstrained mobility, several variants were investigated~\cite{Banu+2012,BDNP14,Garay+1994,Ostrovsky+91,Reischuk+85,Sasaki+2013}: Reischuk~\cite{Reischuk+85} considers that malicious agents are stationary for a given period of time, Ostrovsky and Yung~\cite{Ostrovsky+91} introduce the notion of mobile viruses and define the adversary as an entity that can inject and distribute faults; finally, Garay~\cite{Garay+1994}, and more recently Banu \emph{et al.}~\cite{Banu+2012}, and Sasaki \emph{et al.}~\cite{Sasaki+2013} or Bonnet  \emph{et al.} \cite{BDNP14} consider that processes execute synchronous rounds composed of three phases: \emph{send}, \emph{receive}, and \emph{compute}. Between two consecutive such synchronous rounds, Byzantine agents can move from one node to another. Hence the set of faulty processes at any given time has a bounded size, yet its membership may evolve from one round to the next. The main difference between the aforementioned four works~\cite{Banu+2012,BDNP14,Garay+1994,Sasaki+2013} lies in the knowledge that hosts have about their previous infection by a Byzantine agent. In Garay's model~\cite{Garay+1994}, a host is able to detect its own infection after the Byzantine agent left it. Sasaki \emph{et al.}~\cite{Sasaki+2013} investigate a model where hosts cannot detect when Byzantine agents leave. Finally, Bonnet \emph{et al.}~\cite{BDNP14} considers an intermediate setting where cured hosts remain in \emph{control} on the messages they send (in particular, they send the same message to all destinations, and they do not send obviously fake information, \emph{e.g.} fake id). Those subtle differences on the power of Byzantine agents turns out to have an important impact on the bounds for solving distributed problems.

A first step toward decoupling algorithm rounds from mobile Byzantine moves is due to Bonomi \emph{et al.}~\cite{BDPT16c}. In their solution to the regular register implementation, mobile Byzantine movements are synchronized, but the period of movement is independent to that of algorithm rounds.

Alternatively, \emph{self-stabilization}~\cite{D74j,D00} is a versatile technique to recover from \emph{any number of Byzantine participants}, provided that their malicious actions only spread a \emph{finite} amount of \emph{time}. In more details, starting from an arbitrary global state (that may have been caused by Byzantine participants), a self-stabilizing protocol ensures that problem specification is satisfied again in finite time, without external intervention.\\ 

\noindent\textbf{Register Emulation.} Traditional solutions to  build a Byzantine tolerant storage service (\emph{a.k.a.} register emulation) can be divided into two categories: \emph{replicated state machines} \cite{S90}, and \emph{Byzantine quorum systems} \cite{B00,MR98,MAD02,MAD02-2}. Both approaches are based on the idea that the current state of the storage is replicated among processes, and the main difference lies in the number of replicas that are simultaneously involved in the state maintenance protocol.\\ 

\noindent\textbf{Multi-tolerance.} Extending the effectiveness of self-stabilization to permanent Byzantine faults is a long time challenge in distributed computing. Initial results were mostly negative~\cite{DD05c,DW04j,NA02c} due to the impossibility to distinguish a honest yet incorrectly initialized participant from a truly malicious one. On the positive side, two notable classes of algorithms use some locality property to tolerate Byzantine faults: \emph{space-local} and \emph{time-local} algorithms. Space-local algorithms~\cite{MT07j,NA02c,SOM05c} try to contain the fault (or its effect) as close to its source as possible. This is useful for problems where information from remote nodes is unimportant (such as vertex coloring, link coloring, or dining philosophers). Time-local algorithms~\cite{DMT10ca,DMT11j,DMT11cb} try to limit over time the effect of Byzantine faults. Time-local algorithms presented so far can tolerate the presence of at most a single Byzantine node. Thus, neither approach is suitable to register emulation. 

Recently, several works investigated the emulation of  self-stabilizing or pseudo-stabilizing Byzantine tolerant  SWMR or MWMR registers \cite{AADDPT15,BTP15,BDPR15}. All these works do not consider the complex case of mobile Byzantine faults.

To the best of our knowledge, the problem of tolerating both \emph{arbitrary transient faults and mobile Byzantine faults} has been considered recently only in round-based synchronous systems~\cite{BDP16}. The authors propose optimal \emph{unbounded} self-stabilizing atomic register implementations for \emph{round-based synchronous} systems under the four Mobile Byzantine models described in \cite{Banu+2012,BDNP14,Garay+1994,Sasaki+2013}.\\ 

\noindent\textbf{Our Contribution.} The main contribution of the paper is a protocol $\mathcal{P}_{reg}$ emulating a regular register in a distributed system where both arbitrary transient failures and mobile Byzantine failures can occur. In particular, the proposed solution differs from previous work on round-free register emulation~\cite{BDPT16c,BDPT17} as we add the self-stabilization property. 
%
%
%
In more details, we present a regular register implementation that uses bounded timestamps from the $\mathcal{Z}_{13}$ domain and it is optimal with respect to the number of replicas needed to tolerate $f$ mobile Byzantine agents.
Finally, the convergence time of our solution is upper bounded by $T_{10write()}$, where $T_{10write()}$ is the time needed to execute ten \emph{complete} $write()$ operations.


\section{System Model }\label{sec:systemModel}

We consider a distributed system composed of an arbitrary large set of client processes $\mathcal{C}$ and a set of $n$ server processes $\mathcal{S}=\{s_1, s_2 \dots s_n\}$. Each process in the distributed system (\emph{i.e.}, both servers and clients) is identified by a unique identifier. Servers run a distributed protocol emulating a shared memory abstraction and such protocol is totally transparent to clients (\emph{i.e.}, clients do not know the protocol executed by servers). 
The passage of time is measured by a fictional global clock (\emph{e.g.}, that spans the set of natural integers). 
At each time $t$, each process (either client or server) is characterised by its \emph{internal state}, \emph{i.e.}, by the set of all its local variables and the corresponding values.
No agreement abstraction is assumed to be available at each process (\emph{i.e.} processes are not able to use consensus or total order primitives to agree upon the current values). Moreover, we assume that each process has the same role in the distributed computation (\emph{i.e.}, there is no special process acting as a coordinator).

\noindent{\bf Communication model.} 
Processes communicate through message passing. In particular, we assume that: \emph{(i)} each client $c_i \in \mathcal{C}$ can communicate with every server through a ${\sf broadcast}()$ primitive, \emph{(ii)} each server can communicate with every other server through a ${\sf broadcast}()$ primitive, and \emph{(iii)} each server can communicate with a particular client through a ${\sf send}()$ unicast primitive. We assume that communications are authenticated (\emph{i.e.}, given a message $m$, the identity of its sender cannot be forged) and reliable (\emph{i.e.}, spurious messages are not created and sent messages are neither lost nor duplicated).

\noindent{\bf Timing assumptions.} 
The system is synchronous in the following sense:
\emph{(i)} the processing time of local computations (except for ${\sf wait}()$ statements) is negligible with respect to communication delays and is assumed to be equal to $0$, and \emph{(ii)}  messages take time to travel to their destination processes. 
In particular, concerning point-to-point communications, we assume that if a process sends a message $m$ at time $t$ then it is delivered by time $t+ \delta_p$ (with $\delta_p >0$). Similarly, let $t$ be the time at which a process $p$ invokes the ${\sf broadcast}(m)$ primitive, then there is a constant $\delta_b$ (with $\delta_b \ge \delta_p$) such that all servers have delivered $m$ at time $t+\delta_b$. For the sake of presentation, in the following we consider a unique message delivery delay $\delta$ (equal to $\delta_b \ge \delta_p$), and we assume $\delta$ is known to every process. {\color{black} Moreover, we assume that any process is provided with a physical clock, \emph{i.e.}, non corruptible.}

\noindent{\bf Computation model.} 
Each process of the distributed system executes a distributed protocol $\mathcal{P}_{reg}$ that is composed by a set of distributed algorithms. Each algorithm in $\mathcal{P}_{reg}$ is represented by a finite state automaton and it is composed of a sequence of computation and communication steps. A computation step is represented by the computation executed locally to each process while a communication step is represented by the sending and the delivering events of a message. Computation steps and communication steps are generally called \emph{events}.\\
The computation is \emph{round-free} i.e., the distributed protocol $\mathcal{P}_{reg}$ does not evolve in synchronous rounds and messages can be sent, according to the protocol, at any point in time.\\
Given a process $p_i$ and the protocol $\mathcal{P}_{reg}$, we say that \emph{$p_i$ is correctly executing $\mathcal{P}_{reg}$ in a time interval $[t, t']$} if $p_i$ never deviates from $\mathcal{P}_{reg}$ in $[t, t']$ (i.e., it always follows the automata transitions and never corrupts its local state).

\begin{definition}[Valid State at time $t$]
Let $p_i$ be a process and let $state_{p_i}$ be the state of $p_i$ at some time $t$.
$state_{p_i}$ is said to be valid at time $t$ if it is equal to the state of some fictional process $\bar{p}$ correctly executing $\mathcal{P}_{reg}$ in the interval $[t_0, t]$.
\end{definition}

%
%

\noindent{\bf Failure Model.} 
An arbitrary number of clients may crash while servers are affected by \emph{Mobile Byzantine Failures} i.e., failures are represented by Byzantine agents that are controlled by a powerful external adversary \vir{moving} them from a server to another. We assume that, at any time $t$, at most $f$ mobile Byzantine agents are in system.

In this work we consider the $\Delta$-synchronized and Cured Unaware Model, i.e. $(\Delta S, CUM)$ MBF model, introduced in \cite{BDPT16c} that is suited for round-free computations\footnote{The $(\Delta S, CUM)$ model abstracts distributed systems subjected to proactive rejuvenation \cite{SBAM09} where processes have no self-diagnosis capability.}.
More in details, $(\Delta S, CUM)$ can be specified as follows: the external adversary moves all the $f$ mobile Byzantine agents at the same time $t$ and movements happen periodically (i.e., movements happen at time $t_0+\Delta$, $t_0+2\Delta$, . . . , $t_0+i\Delta$, with $i \in \mathbb{N}$) and at any time $t$, no process is aware about its failure state (i.e., processes do not know if and when they have been affected from a Byzantine agent) but it is aware about the time at which mobile Byzantine agents move.

Let us note that when we are considering Mobile Byzantine agents no process is guaranteed to be in the same failure state for ever. Processes, in fact, may change their state between \emph{correct} and \emph{faulty} infinitely often.
As a consequence, it is fundamental to re-define the notion of correct and faulty process as follows:

\begin{definition}[Correct process at time $t$]
A process is said to be \emph{correct at time $t$} if (i) it is correctly executing its protocol $\mathcal{P}$ and (ii) its state is a valid state at time $t$.
We will denote as $Co(t)$ the set of correct processes at time $t$ while, given a time interval $[t, t']$, we will denote as $Co ([t, t'])$ the set of all the processes that are correct during the whole interval $[t, t']$ (i.e., $Co ([t, t'])= \bigcap_{\tau ~\in ~[t, t']} Co(\tau)$).
\end{definition}

\begin{definition}[Faulty process at time $t$]
A process is said to be \emph{faulty at time $t$} if it is controlled by a mobile Byzantine agent and it is not executing correctly its protocol $\mathcal{P}$ (i.e., it is behaving arbitrarily). 
We will denote as $B(t)$ the set of faulty processes at time $t$ while, given a time interval $[t, t']$, we will denote as $B ([t, t'])$ the set of all the processes that are faulty during the whole interval $[t, t']$ (i.e., $B ([t, t'])= \bigcap_{\tau ~\in ~[t, t']} B(\tau)$).
\end{definition}

\begin{definition}[Cured process at time $t$]
A process is said to be \emph{cured at time $t$} if (i) it is correctly executing its protocol $\mathcal{P}$ and (ii) its state is not a valid state at time $t$.
We will denote as $Cu(t)$ the set of cured processes at time $t$ while, given a time interval $[t, t']$, we will denote as $Cu ([t, t'])$ the set of all the processes that are cured during the whole interval $[t, t']$ (i.e., $Cu ([t, t'])= \bigcap_{\tau ~\in ~[t, t']} Cu(\tau)$).
\end{definition}

As in the case of round-based MBF models  \cite{Banu+2012,BDNP14,Garay+95+ORA,Garay+1994,Sasaki+2013}, we assume that any process has access to a tamper-proof memory storing the correct protocol code.

Let us stress that even though at any time $t$, at most $f$ servers can be controlled by Byzantine agents, during the system life time, all servers may be affected by a Byzantine agent (\emph{i.e.}, none of the servers is guaranteed to be correct forever). 

Processes may also suffer from {\em transient} failures, \emph{i.e.}, local variables of any process (clients and servers) can be arbitrarily modified~\cite{D00}. 
It is nevertheless assumed that transient failures are quiescent, \emph{i.e.}, there exists a time $\tau_{no\_tr}$ (which is unknown to the processes) after which no new transient failures happens.

\section{Self-Stabilizing Regular Register Specification}\label{sec:Problem}

A register is a shared variable  accessed by a set of processes, \emph{i.e.} clients,  through two operations, namely ${\sf read}()$ and ${\sf write}()$. Informally, the ${\sf  write}()$ operation updates the value stored in the shared variable while  the ${\sf read}()$ obtains the value contained in the variable (\emph{i.e.} the last written value). 
In distributed settings, every operation issued on a register is, generally, not instantaneous and it  can  be characterized  by  two events  occurring  at  its boundary:  an \emph{invocation} event and a \emph{reply} event.
An operation $op$  is \emph{complete} if both the  invocation event and the reply event occur (\emph{i.e.} the  process executing the operation does not crash between the invocation and the reply).
Contrary, an operation $op$ is said to be \emph{failed} if it is invoked by a process that crashes before the reply event occurs. According to these time instants, it is possible to state when two operations are concurrent with respect to the real time execution.
Given two operations  $op$ and $op'$, their  invocation event  times  ($t_{B}(op)$ and $t_B(op')$) and their reply event times ($t_E(op)$ and $t_E(op')$),  we say that $op$ \emph{precedes} $op'$ ($op \prec op'$) iff $t_E(op) < t_B(op')$. If $op$ does not precede $op'$ and $op'$  does not  precede $op$,  then $op$  and $op'$   are \emph{concurrent} ($op||op'$). 
Given a  ${\sf write}(v)$ operation,  the value $v$  is said to  be written when the operation is complete.\\ 
We assume that locally any client never performs ${\sf read()}$ and ${\sf write()}$ operations concurrently (\emph{i.e.}, for any given client $c_i$, the set of operations executed by $c_i$ is totally ordered). We also assume that initially the register stores a default value $\bot$ written by a fictional ${\sf write}(\bot)$ operation happening instantaneously at time $t_0$. 
In case of concurrency while  accessing the shared variable, the meaning of \emph{last written  value} becomes ambiguous.  Depending  on the  semantics of the operations, three types of register  have   been   defined   by  Lamport \cite{L86}:   \emph{safe}, \emph{regular} and \emph{atomic}. 

In this paper, we consider a Self-Stabilizing Single-Writer/ Multi-Reader (SWMR) regular register, \emph{i.e.}, an extension of Lamport's regular register that considers transitory failures. 

The Self-Stabilizing Single-Writer/Multi-Reader (SWMR) register is specified as follow:

\begin{itemize}
	\item ${\sf ss-Termination}$: Any operation invoked on the register by a non-crashed process eventually terminates. 
	
	\item ${\sf ss-Validity}$: There exists a time $t_{stab}$ such that each ${\sf read}()$ operation invoked at time $t > t_{stab}$ returns the last value written before its invocation, or a value written by a ${\sf write}()$ operation concurrent with it.
\end{itemize}

\section{A Self-Stabilizing Regular Register Implementation}\label{sec:algorithm}

In this Section we propose a protocol $\mathcal{P}_{reg}$ implementing a self-stabilizing SWMR regular register in the $(\Delta S, CUM)$ Mobile Byzantine Failure model. Such an algorithm copes with the $(\Delta S, CUM)$ model following the same approach as in \cite{BDPT17}, which is improved with the bounded timestamps in order to design a self-stabilizing algorithm.    

We implemented ${\sf read}()$ and ${\sf write}()$ operations following the classical quorum-based approach (like in the ABD protocol \cite{ABD95}) and exploiting the synchrony of the system to guarantee their termination. 
Informally, when the writer client wants to write, it simply propagates the new value to servers that update the value of the register while, when a reader client wants to read, it asks for the current value of the register and waits for replies: after $3 \delta$ time \vir{enough}\footnote{The exact number of replies is provided in Table \ref{tab:summaryCUM} depending on the relationship between $\Delta$ and $\delta$.} replies have been received and a value is selected and returned (the reason why a ${\sf read}()$ operation lasts for $3\delta$ is explained in the following).

In order to do that, the ${\sf maintenance}()$ operation must guarantee that there always exists a sufficient number of servers storing a valid value for the register. Thus, its aim is threefold: (i) ensuring that cured servers get a valid value at the end of ${\sf maintenance}()$, (ii) possible concurrent written values are always taken into account by cured servers running ${\sf maintenance}()$ and (iii) correct servers do not overwrite their correct value with a non-valid one.

Each server $s_i$ stores three pairs $\langle value, timestamp \rangle$ corresponding to the last three written values and periodically (when Byzantine agents move at every $T_i=t_0 + i\Delta$, with $i \in \mathbb{N}$) executes the ${\sf maintenance}()$ operation.


The basic idea is to keep separated information that can be trusted (e.g., values received by the writer client or values sent from \vir{enough} processes) from those that are untrusted (e.g., values stored locally that can be compromised) and to decide the current state accordingly.

To this aim, ${\sf maintenance}()$ makes use of three fundamental set variables: (i) $V_i$ stores the  knowledge of $s_i$  at the beginning of each ${\sf maintenance}()$  operation and contains the last three values of the register and the corresponding sequence numbers (untrusted information), (ii) $V_{safe}$ (emptied at the  beginning of each the ${\sf maintenance}()$  operation) is used to collect values selected among those sent through echoes by other servers (trusted information due to the presence of \vir{enough} correct servers) and (iii) $W_i$ contains values and the corresponding timestamps concurrently received by the writer (untrusted information as it can be potentially compromised by the Byzantine agent before it leaves the server).

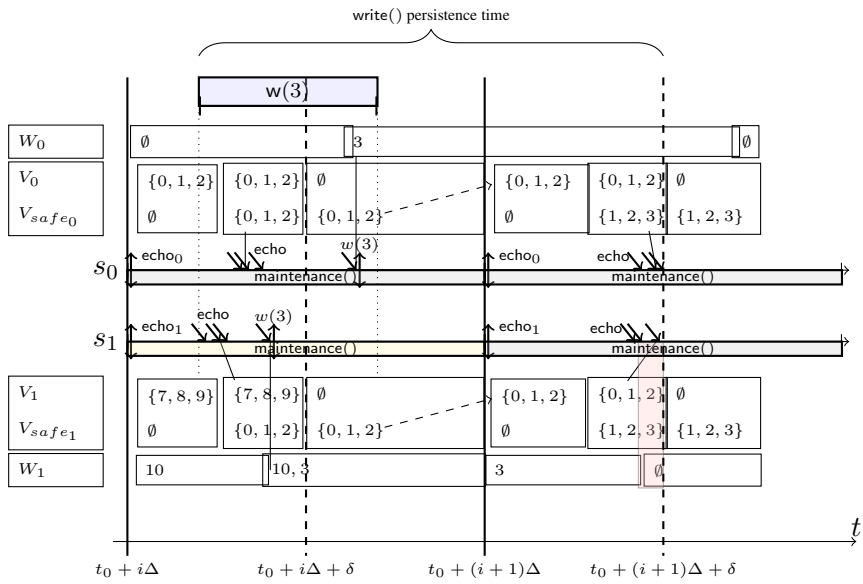
\begin{figure}
	\centering
\begin{tikzpicture}[y=-1cm, scale=0.95]
 
 	\def \lenght {10.1}
 	\def \n {1} 
 	
 	\def \deltaGrande {5}
 	\def \deltaPiccolo {2.5}
 	\def \gammaCuring {1}
 	
 	\processes
	\foreach \t in {0,...,1}{
		\foreach \x in {0,\deltaGrande,...,5}{
			\draw [|-|] (\x,\t)--(\x+\deltaGrande, \t);
		}
	}
	\node[] () at (-1,-1.8) [draw, text width=1cm]{\tiny $W_0$};
	\node[] () at (-1,-1) [draw, text width=1cm]{\tiny $V_0$ \\ $V_{safe_0}$};
	\node[] () at (.7,-1) [draw, text width=.8cm]{\tiny $\{ 0,1,2\}$ \\ $\emptyset$};
	\node[] () at (1.6,-1.8) [draw, text width=2.7cm]{\tiny $\emptyset$};
	\node[] () at (1.9,-1) [draw, text width=.8cm]{\tiny $\{ 0,1,2\}$ \\ $\{ 0,1,2\}$};	\draw[] (1.65,0)--(1.65,-.54);
	\draw[] (3.2,0)--(3.2,-1.6);
	\node[] () at (3.75,-1) [draw, text width=2.1cm]{\tiny $\emptyset$ \\ $\{ 0,1,2\}$};
	\node[] () at (5.8,-1.8) [draw, text width=5cm]{\tiny $3$};
	\draw[->, dashed] (3.6,-.8) -- (5.1,-1.2);

	\node[] () at (5.8,-1) [draw, text width=1cm]{\tiny $\{ 0,1,2\}$ \\ $\emptyset$};
	\node[] () at (7,-1) [draw, text width=.8cm]{\tiny $\{ 0,1,2\}$ \\ $\{1,2,3\}$}; \draw[] (7.4,1)--(7,1.55);
	\node[] () at (8.2,-1) [draw, text width=1cm]{\tiny $\emptyset$ \\ $\{ 1,2,3\}$};
	\node[] () at (8.65,-1.8) [draw, text width=.1cm]{\tiny $\emptyset$};	
	
	\node[] () at (-1,2.8) [draw, text width=1cm]{\tiny $W_1$};
	\node[] () at (-1,2) [draw, text width=1cm]{\tiny $V_1$ \\ $V_{safe_1}$};
	\node[] () at (.7,2) [draw, text width=.8cm]{\tiny $\{ 7,8,9\}$ \\ $\emptyset$};
	\node[] () at (1.05,2.8) [draw, text width=1.5cm]{\tiny $10$};
	\node[] () at (1.9,2) [draw, text width=.8cm]{\tiny $\{ 7,8,9\}$ \\ $\{ 0,1,2\}$}; \draw[] (1.3,1)--(1.5,1.54);
	\draw[] (2,1)--(2,2.8);
	\node[] () at (3.45,2.8) [draw, text width=2.7cm]{\tiny $10,3$};
	\node[] () at (6.1,2.8) [draw, text width=1.8cm]{\tiny $3$};	
	\node[] () at (3.75,2) [draw, text width=2.1cm]{\tiny $\emptyset$ \\ $\{ 0,1,2\}$};
	\draw[->, dashed] (3.6,2.2) -- (5.1,1.8);
	\node[] () at (5.75,2) [draw, text width=1cm]{\tiny $\{ 0,1,2\}$ \\ $\emptyset$};
	\node[] () at (7,2) [draw, text width=.8cm]{\tiny $\{ 0,1,2\}$ \\ $\{1,2,3\}$}; \draw[] (7.4,0)--(7.3,-.55);
	\node[] () at (8.2,2) [draw, text width=1cm]{\tiny $\emptyset$ \\ $\{ 1,2,3\}$};
	\node[] () at (8.05,2.8) [draw, text width=1.3cm]{\tiny $\emptyset$};		
	
	\bc{.05}{0}{black} \node[] () at (.5, -.2) {\tiny ${\sf echo}_0$};
	\bc{.05}{1}{black} \node[] () at (.5, .8) {\tiny ${\sf echo}_1$};
	\req{1.5}{0}{black} \req{1.4}{0}{black} \req{1.7}{0}{black} \node[] () at (2, -.3) {\tiny ${\sf echo}$};
	\req{1.1}{1}{black} \req{.9}{1}{black} \req{1.2}{1}{black}	\node[] () at (1.2, .6) {\tiny ${\sf echo}$};
	
	\req{3}{0}{black} \bc{3.25}{0}{black} \node[] () at (3.25, -.35) {\tiny $w(3)$};
	\req{1.8}{1}{black} \bc{2.05}{1}{black} \node[] () at (2.05, .65) {\tiny $w(3)$};

	\bc{5.05}{0}{black} \node[] () at (5.5, -.2) {\tiny ${\sf echo}_0$};
	\bc{5.05}{1}{black} \node[] () at (5.5, .8) {\tiny ${\sf echo}_1$};
	\req{7.2}{0}{black} \req{7}{0}{black}  \req{7.3}{0}{black}  \node[] () at (6.8, -.2) {\tiny ${\sf echo}$};
	\req{7.25}{1}{black} \req{6.9}{1}{black} \req{7}{1}{black}	\node[] () at (6.7, .8) {\tiny ${\sf echo}$};

\draw[fill=red!20, opacity=0.3]	(7.15,1) rectangle (7.5,3.05);

\draw[fill=gray!50, opacity=0.2] (0,0) rectangle (5,.2);	\draw[draw, thick] (0,0) rectangle (5,.2);
\node[]() at (2.5,.1) {\tiny \op{maintenance}};
\draw[fill=gray!50, opacity=0.2] (5,0) rectangle (10,.2);	\draw[draw, thick] (5,0) rectangle (10,.2);
\node[]() at (7.5,.1) {\tiny \op{maintenance}};

\draw[fill=yellow!50, opacity=0.2] (0,1) rectangle (5,1.2);	\draw[draw, thick] (0,1) rectangle (5,1.2);
\node[]() at (2.5,1.1) {\tiny \op{maintenance}};
\draw[fill=gray!50, opacity=0.2] (5,1) rectangle (10,1.2);	\draw[draw, thick] (5,1) rectangle (10,1.2);
\node[]() at (7.5,1.1) {\tiny \op{maintenance}};
	
\lineaVerticaleLabelBasso{0}{-2.7}{\tiny $t_0+i\Delta$}{thick}{4}
\lineaVerticaleLabelBasso{2.5}{-2.7}{\tiny $t_0+i\Delta+\delta$}{dashed, thick}{4}

\lineaVerticaleLabelBasso{5}{-2.7}{\tiny $t_0+(i+1)\Delta$}{thick}{4}	
\lineaVerticaleLabelBasso{7.5}{-2.7}{\tiny $t_0+(i+1)\Delta+\delta$}{dashed, thick}{4}

\draw[->] (-.2,3.8)--(10.2,3.8); \node[] () at (10.2,3.6) {$t$};

\draw[fill=blue!20, opacity=0.3]	(1,-2.3) rectangle (3.5,-2.7);
\draw[draw, thick]	(1,-2.3) rectangle (3.5,-2.7);	
\draw [decorate,decoration={brace,amplitude=10pt},xshift=0,yshift=0]
(1,-3) -- (7.5,-3) node [black,midway,yshift=15] 
{\tiny \op{write} persistence time};
\scritturaValoreAltezza{1}{3}{-2.3}
	
\end{tikzpicture}
\caption{Example of a partial run for a correct server $s_0$ and a cured server $s_1$ with $\Delta=2\delta$. For the sake of simplicity we report only timestamps instead of the pair $\langle value, timestamp\rangle$.}\label{fig:detailedRun}
\end{figure}

As an example, consider the execution of the $i$-th ${\sf maintenance}()$ operation starting at time $t_0+i\Delta$ shown in Figure \ref{fig:detailedRun} for the two servers $s_0$ and $s_1$ that are respectively correct and cured.

When ${\sf maintenance}()$ starts, every server $s_i$ echoes the relevant information stored locally (i.e., list of pending \op{read} operations and the sets $V_i$ and $W_i$). 
Such information are then collected by any server $s_j$ and can be used (based on the number of occurrence of each pair $\langle$value, sequence number$\rangle$) to update the set $V_{safe}$.
Let us note that, due to the synchrony of the system, after $\delta$ time units (i.e., at time $t_0+i\Delta+\delta$), $s_i$ collected at least all the values sent by every correct and cured server and it is able to decide and update its local variables. Thus, it selects the values occurring \vir{enough times} (see footnote 2)
from echoes, updates $V_{safe}$ and empties $V_i$. 

In Figure \ref{fig:detailedRun}, it is possible to see that, at time $t_0+i\Delta+\delta$, $s_0$ basically does not update its information while $s_1$ is able to update its $V_{safe}$ set consistently with $s_0$ using the values gathered through echoes.

However, the \op{maintenance} operation is not yet terminated as it could happen that a ${\sf write}()$ operation is running concurrently and the concurrently written value may not yet be in $V_i$ and in $V_{safe}$. In order to manage this case, every time that a value is written, it is also relayed to all servers.
In addition, in order to avoid to overwrite values just written with those selected from the \op{maintenance} operation, concurrently written values are temporarily stored in $W_i$ with an associated timer (i.e., like a time-to-leave) set to $2\delta$.  

The timer is set in such a way that each value in $W_i$ remains stored long enough to ensure its propagation to all servers and guarantees that written values will eventually appear in every non-faulty $V_{safe}$ set (e.g., the value $3$ in Figure \ref{fig:detailedRun}). 
At the same time, the $2\delta$ period is not long enough to allow mobile Byzantine agents to leverage the propagation of corrupted values to force reader clients to return a bad value (e.g., the value $10$ left by the mobile Byzantine agent in $W_1$ at the beginning of the $i$-th ${\sf maintenance}()$ operation).  We call the time necessary for a value to be present in $V_{safe}$ as the write persistence time. 




Note that, depending on the relationship between $\Delta$ and $\delta$, it may happen that a ${\sf maintenance}()$ operation is triggered while the previous one is not yet terminated. This is not an issue as the set $V_i$ is updated before the second ${\sf maintenance}()$ operation starts and $W_i$ is the only set that is not reset between the two \op{maintenance} operations and it prevents values to be lost having a time-to-live of $2\delta$ which is enough to propagate it.

Finally, concurrently with the ${\sf maintenance}()$ and \op{write} operations, servers may need to answer also to clients that are currently reading. In order to preserve the validity of ${\sf read}()$ operations and in order to cope with possible corrupted values stored by $s_i$ just before the mobile Byzantine agent left, $s_i$ replies with all the values it is storing (i.e., providing $V_i$, $V_{safe}$ and $W_i$). 
Note that, given the update mechanism of local variables (designed to keep separated trusted information from untrusted ones), there could be a fraction of time where the last written value is removed from $W_i$ (as its timer is expired) and it is not yet inserted in $V_i$ and in $V_{safe}$ (as the corresponding propagation message is still traveling - cfr. the red zone in Figure \ref{fig:detailedRun}). 
To cope with this issue, the ${\sf read}()$ operation lasts $3\delta$ time i.e., an extra waiting period is added for the collection of replies to guarantee that values are not lost.\\




In order to stabilise in a finite and known period and manage transient failures, $\mathcal{P}_{reg}$ employs bounded timestamps. 
It is important to note that timestamps are necessary in the $(\Delta S, CUM)$ model as, during the ${\sf maintenance}()$, servers must be able to distinguish new and old values in order to guarantee that a new value possibly received by the writer is not overwritten by the \op{maintenance} operation.
In the following we will explain why using bounded timestamps guarantees a  finite and known stabilisation period.

Let us note that, in order to stabilise, at least one \op{write} operation must be executed after time $\tau_{no\_tr}$. However, due to the fact that this operation is the first one after $\tau_{no\_tr}$, if the domain of timestamps is unbounded (e.g., the domain of natural numbers $\mathbb{N}$ as in \cite{BDP16,BDPT16c,BDPT17}), it could happen that the timestamp used by the writer is way much smaller than those stored locally by servers. This means that such an operation will be ignored and the same will happen until the writer timestamp will reach those stored by servers making the stabilisation period unknown.


We use timestamps in the domain $\mathcal{Z}_m$, with $m=13$. Each written value is represented  as $\ang{val}{sn}$ where $val$ is the content and $sn$ the corresponding sequence number, $sn \in Z_m=\{0,1,\dots,m-1\}$. Let us define two operations on such values: addition: $+_m:Z_m \times Z_m \rightarrow Z_m, a +_m b = (a+b) \Mod{m}$; and subtraction: $-_m:Z_m \times Z_m \rightarrow Z_m, a -_m b = a +_m (-b)$.  Note that $(-b)$ is the opposite of $b$. That is, the number that added to $b$ gives $0$ as result, \emph{i.e.}, $b +_m (-b)=0$. 

\begin{table*}[t]
\centering
	\scriptsize
	\begin{tabular}{| c | c | c | c |}
		\hline
		\\[-1em]
		$k=\lceil \frac{3\delta}{\Delta}\rceil$& ${{n_{CUM}} \ge (2k+2)f+1}$ &  $ {\# reply_{CUM} \geq 2kf+1}$ & ${\# echo_{CUM} \geq kf+1}$  \\
		\hline
		$\Delta = \delta, k=3$  & $8f+1$ &  $6f+1$ & $3f+1$ \\
		\hline
		$\Delta = 2\delta, k=2$  & $6f+1$ &  $4f+1$ & $2f+1$ \\
		\hline		
	\end{tabular}
	\caption{Parameters for $\mathcal{P}_{Rreg}$ Protocol.}\label{tab:summaryCUM}
\end{table*}

\begin{figure*}
	\centering
\begin{minipage}{.5\textwidth}
	\begin{tikzpicture}[y=-1cm, yscale=.8]
		\def \lenght {3}
		\def \n {2} 
		
		\def \deltaGrande {.5}
		\def \deltaPiccolo {.7}
		\def \gammaCuring {1}

		\processes
		\foreach \t in {0,...,2}{
			\foreach \x in {0,.7, 1.4,...,2.1}{
				\draw [|-|] (\x,\t)--(\x+.7, \t);
			}
		}

		
\draw [decorate,decoration={brace,amplitude=5pt},xshift=0,yshift=0]
			(.5,-1.4) -- (2.1,-1.4) node [black,midway,yshift=10] 
			{\tiny ${\sf w}(3)$ persistence time};
		\scritturaValore{.5}{3}
		\scritturaValore{1.2}{4}
		\scritturaValore{1.9}{5}
		
		\node[] () at (2.1,-.2) {\scriptsize \color{red} (a)};
	\end{tikzpicture}
	\hspace{.2cm}
  \begin{tikzpicture}[y=-1cm, rotate=90, scale=.5]
  	\def \angolo {27.6}
    \draw[black, thick] (0,0) circle (2);    
    \foreach \a in {0, \angolo,...,360}{
    	\draw[thick,black] (0, 0) -- (\a:2.3);
    }
    \foreach \b in {0,...,12}{
   		\draw (\angolo*\b: 3) node {\scriptsize \b};
    
    }
    \draw[fill, red] (0: 2) circle (.1cm);
    \draw[fill, red] (\angolo*1: 2) circle (.1cm); 
    \draw[fill, red] (\angolo*2: 2) circle (.1cm);
    \draw[fill, red] (\angolo*5: 2) circle (.1cm);  
    \draw[fill=black] (0,0) circle(0.7mm);
    \draw [-latex] (-1,.2) arc [start angle=200, end angle=-130, x radius=1cm, y radius=1cm];
    \draw [-latex, red, dotted] (-1.5,-2.2)  arc [start angle=150, end angle=10, x radius=1.5cm, y radius=.7cm];
    \draw [-latex] (1.5,2.2) arc [start angle=150, end angle=0, x radius=-1.5cm, y radius=-.7cm];
  \end{tikzpicture}
\end{minipage}%
\begin{minipage}{.5\textwidth}
  		\begin{tikzpicture}[y=-1cm, yscale=.8]
  			\def \lenght {3.8}
  			\def \n {2} 
  			
  			\def \deltaGrande {.5}
  			\def \deltaPiccolo {.7}
  			\def \gammaCuring {1}

  			\processes
  			\foreach \t in {0,...,2}{
  				\foreach \x in {0,.7, 1.4,...,2.8}{
  					\draw [|-|] (\x,\t)--(\x+.7, \t);
  				}
  			}

  			\scritturaValore{.2}{3}
  			\scritturaValore{.9}{4}
  			\scritturaValore{1.6}{5}
  			\scritturaValore{2.3}{6}
  			\scritturaValore{3}{7}
  			\letturaLunga{1}{3}
  			\node[] () at (2,-.7) {\scriptsize$r()$};
  			
  		\end{tikzpicture}
  		\hspace{.2cm}
  \begin{tikzpicture}[y=-1cm, rotate=90, scale=.5]
 	\def \angolo {27.6}
    \draw[black, thick] (0,0) circle (2);    
    \foreach \a in {0, \angolo,...,360}{
    	\draw[thick,black] (0, 0) -- (\a:2.3);
    }
    \foreach \b in {0,...,12}{
   		\draw (\angolo*\b: 3) node {\scriptsize \b};
    
    } 
    \draw[fill, red] (0: 2) circle (.1cm);
    \draw[fill, red] (\angolo*1: 2) circle (.1cm); 
    \draw[fill, red] (\angolo*2: 2) circle (.1cm); 
    \draw[fill, red] (\angolo*3: 2) circle (.2cm);
    \draw[fill, red] (\angolo*4: 2) circle (.1cm);
    \draw[fill, red] (\angolo*5: 2) circle (.1cm);
    \draw[fill, red] (\angolo*6: 2) circle (.1cm); 
    \draw[fill, red] (\angolo*7: 2) circle (.1cm);  
    \draw[fill=black] (0,0) circle(0.7mm);
    \draw [-latex] (-1,.2) arc [start angle=200, end angle=-130, x radius=1cm, y radius=1cm];
    \draw [-latex, red, dotted] (-1.5,-2.2)  arc [start angle=150, end angle=10, x radius=1.5cm, y radius=.7cm];
    \draw [-latex] (1.5,2.2) arc [start angle=150, end angle=0, x radius=-1.5cm, y radius=-.7cm];
      \end{tikzpicture}

\end{minipage}
\caption{Runs for $\Delta=\delta$. The small vertical lines are the points where the \op{maintenance} operations begin. For simplicity we represent values with their timestamp and we consider only correct servers $s_i$ that store $V_i=\{0,1,2\}$.}\label{fig:examples}
\end{figure*}
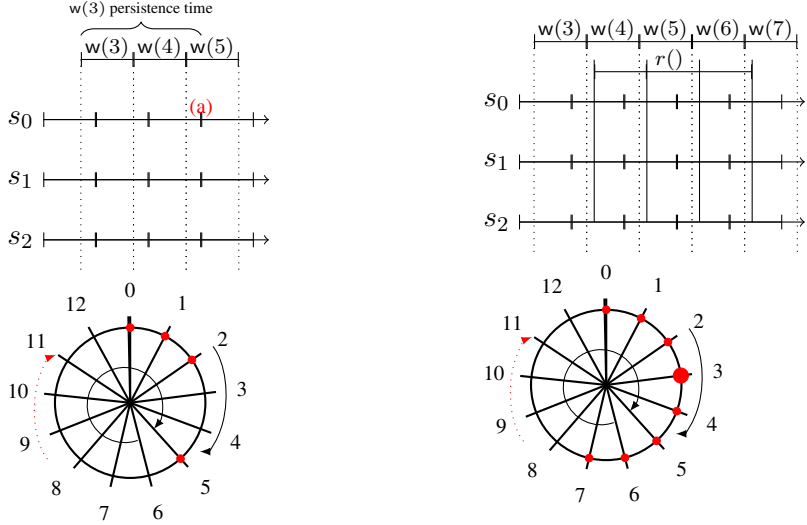

Two scenarios are depicted in Figure \ref{fig:examples} to characterize how many different values clients and servers may have to manage (and thus uniquely order) at the same time. We consider a sequence of \op{write} operations and then what happens if a \op{read} operation is concurrent with a sequence of \op{write} operations. In the first case, just before the time instant marked as (a) $s_0$ could be ready to store values $0,1,2$ and $5$ that need to be ordered. In the meantime values $3$ and $4$ are still echoed. In any case the timestamps range that a server can manage at the same time is from $0$ to $5$, more in general $6$ subsequent timestamps. $3$ values are stored in $V_i$ and $3$ values come from the subsequent \op{write} operations. At time (a), in Figure \ref{fig:examples}, $3$ takes the place of $0$, which is discharged. Let us consider the most distant values, $0$ and $5$. There are two ways to order them, either $0$ precedes $5$ or $5$ precedes $0$. But the second one is impossible since in that case there could be $7$ timestamps around at the same time. In the second scenario, concurrently to a sequence of \op{write} operations there is a \op{read} operation. In this case we have to consider all values that could be returned to the client. In this case, values from $0$ to $7$ (thus at most at distance $7$) and we notice that the last written value, $3$, is always returned. Thus a client may have to order the following values $0,3,7$. There are three possibilities: (i) $0,3,7$, (ii) $3,7,0$, or (iii) $7,0,3$. In cases (ii) and (iii) we have $0-_{13}3=10$ and $3-_{13}7=9$ respectively, both of them greater than $7$. Thus the only possible order the case (i). 

The pseudo-code for $\mathcal{P}_{reg}$ is shown in Figures \ref{fig:stateMaintenanceProtocolCUM} - \ref{fig:readProtocolCUM}. 

\noindent{\bf Local variables at client $c_i$.} Each client $c_i$ maintains two sets $reply_i$ that is used during the ${\sf read}()$ operation to collect the three tuples $\langle j, \langle v, sn \rangle \rangle$ sent back from servers. 
Additionally, if $c_i$ is the writer, it maintains a local sequence number $csn$ that is incremented, respect to the $\mathcal{Z}_{13}$ arithmetic, each time it invokes a ${\sf write}()$ operation, which is timestamped with such sequence number.\\

\noindent{\bf Local variables at server $s_i$.} Each server $s_i$ maintains the following local variables: 
\begin{itemize}
	
	
	
	\item $V_i$: a set containing $3$ tuples $\langle v, sn \rangle$, where $v$ is a value and $sn$ the corresponding sequence number.
	\item$V_{safe_i}$: this set has the same characteristic as $V_i$, and is populated by the function ${\sf insert}(V_{safe_i}, \langle v_k, sn_k \rangle)$.
	\item $W_i$: a set where $s_i$ stores values coming directly from the writer, associating to it a timer, $\langle v, sn, timer \rangle$. When the timer expires, the associated value is deleted.
	\item $echo\_vals_i$ and $echo\_read_i$: two sets used to collect information propagated through {\sc echo} messages. The former set stores tuple $\langle v, sn \rangle_j $ whilst the latter set contains identifiers of concurrently reading clients in order to notify cured servers and expedite termination of ${\sf read}()$ operations.
	\item$pending\_read_i$: set variable used to collect identifiers of the clients that are currently reading. Notice, for simplicity we do not explicitly manage the values discharge from such set since it has no impact on the protocol correctness.
	
\end{itemize}

In order to simplify the code of the algorithm, let us define the following functions:

\begin{itemize}
	\item ${\sf select\_pairs} (echo\_vals_i)$: this function takes as input the set $echo\_vals_i$ and returns tuples $\langle v, sn \rangle$, such that there exist at least $\# echo_{CUM}$ occurrences in $echo\_vals_i$ of such tuple (ignoring the Timer value, if present). 
	\item ${\sf insert}(V_{safe_i}, \langle v_k, sn_k \rangle)$: this function inserts $\langle v_k, sn_k \rangle$ in $V_{safe}$ according with the incremental order and if there are more than $3$ values then the oldest one is discarded. In case it is not possible to establish an unique order among the elements in the set then $V_{safe_i}$ is reset (this may happen due to transient failures).
	\item ${\sf select\_value}(reply_i)$: this function returns the newest pair $\langle v, sn \rangle$ occurring at least $\#reply_{CUM}$ times in $reply_i$  (ignoring the Timer value, if present).
	\item${\sf checkOrderAndTrunc}(V_{safe_i})$: this function checks if it is possible to uniquely order the elements in $V_{safe_i}$ with respect to the timestamps. If yes, the $3$ newest element are kept, the others are discharged. If it is not possible to uniquely establish an order for each pair of elements then all the elements are discharged.
	\item${\sf checkOrder}(V_{safe_i})$: this function checks if it is possible to establish an unique order for each couple of elements in $V_{safe_i}$. If not, $V_{safe_i}$ is emptied.
	\item ${\sf conCut}(V_i, V_{safe_i},W_i)$: this function takes as input three $3$ dimension ordered sets and returns another $3$ dimension ordered set. The returned set is composed by the concatenation of $V_{safe_i} \circ V_i \circ W_i$, without duplicates, truncated after the first $3$ newest values (with respect to the timestamp). e.g., $d=3$, $V_i=\{\ang{v_a}{1},\ang{v_b}{2}, \ang{v_c}{3}\}$ and $V_{safe_i}=\{\ang{v_b}{2}, \ang{v_d}{4}, \ang{v_f}{5}\}$ and $W_i=\emptyset$, then the returned set is $\{\ang{v_c}{3}, \ang{v_d}{4}, \ang{v_f}{5}\}$. If is it not possible to establish and order  in one of those sets because of transient failures then the result is $\bot$.
	\item ${\sf checkTimer}(W_i)$: this function removes from $W_i$ all the values whose associated timer is $0$ or strictly greater than $2\delta$.
	
\end{itemize}

\noindent{\bf The ${\sf maintenance}()$ operation.} Such operation is executed by servers periodically at any time $T_i=t_0+i\Delta$. Each server first stores the content of $V_{safe_i}$ in $V_i$ and all $V_{safe_i}$ and $echo\_vals_i$ sets are reset. Each server broadcasts an {\sc echo} message with the content of $V_i$, $W_i$ and the $pending\_read_i$ set. When there is a value in the $echo\_vals_i$ set that occurs at least $\#echo_{CUM}$ times, $s_i$ tries to update $V_{safe_i}$ set by invoking ${\sf insert}$ on the value returned by the ${\sf select\_pairs} (echo\_vals_i)$ function. To conclude, after $\delta$ time since the beginning of the operation, the $V_i$ set is reset. Informally speaking, during the \op{maintenance} operation $V_{safe_i}$ is filled with safe values, then the content in $V_i$ is not longer necessary. Notice that the content of $W_i$ is continuously monitored so that expired values are removed.\\

\begin{figure*}[!]
\centering
\fbox{
\begin{minipage}{0.4\textwidth}
\scriptsize
\resetline
\begin{tabbing}
aaaA\=aA\=aA\=aaaA\kill

${\sf init}()$ :\\

\line{}\> {\bf trigger} ${\sf maintenance}()$; ${\sf checkTimer}(W_i)$; ${\sf select} (echo\_vals_i)$;\\

%
%

%

%
%
%
%

~------------------------------------------------------------------------------------------------------\\

{\bf operation} ${\sf maintenance}()$ {\bf executed every} $T_i=t_0+\Delta_i$ :\\

\line{M-CUM-00} \> ${\sf checkOrderAndTrunc}(V_{safe_i})$;\\


\line{M-CUM-01} \> $echo\_vals_i\leftarrow \emptyset$; $V_i\leftarrow V_{safe_i}$; $V_{safe} \leftarrow \emptyset$; \\

%
%
%

\line{M-CUM-06} \> ${\sf broadcast}$ {\sc echo}$(i,  V_i \cup W_i, pending\_read_i)$;\\
\line{M-CUM-06a} \>  {\bf wait}$(\delta)$;\\

\line{M-CUM-06b} \> $V_i \leftarrow \emptyset$;\\

~-------------------------------------------------------------------------------------------------------------\\

{\bf when} \=  {\sc echo} $(j, S, pr)$ \= is  ${\sf received}$: \\ 

\line{M-CUM-12} \> {\bf for each} \= $(\langle v, sn \rangle_j \in S)$\\

\line{M-CUM-13} \>\> $echo\_vals_i \leftarrow echo\_vals_i \cup \langle v, sn \rangle_j$;\\

\line{M-CUM-14} \>  {\bf endFor}\\
\line{M-CUM-15} \> $echo\_read_i \leftarrow echo\_read_i \cup pr$;\\

~------------------------------------------------------------------------------------------------------\\

{\bf function}  ${\sf select} (echo\_vals_i)$:\\




\line{} \> {\bf while}(\sc true):\\

\line{M-CUM-13} \>\> {\bf if} \= ${\sf select\_pairs} (echo\_vals_i)\neq \bot$;\\

\line{M-CUM-13} \>\>\> $\ang{v_k}{sn_k} \leftarrow {\sf select\_pairs} (echo\_vals_i)$;\\


\line{M-CUM-15}\>\>\> ${\sf insert}(V_{safe_i},  \ang{v_k}{sn_k})$;\\






\line{M-CUM-10} \>\>\> ${\sf send}$ {\sc reply} $(i, {\sf conCut}(V_i,V_{safe_i},W_i)))$ to $c_j$;\\

\line{M-CUM-11} \>\>{\bf endIF}\\

\line{M-CUM-11} \>{\bf endWhile}\\

\end{tabbing}
\normalsize
\end{minipage}
}
\caption{$\mathcal{A}_{M}$ algorithm implementing the ${\sf maintenance}()$ operation (code for server $s_i$) in the $(\Delta S, CUM)$ model with bounded timestamp.}
\label{fig:stateMaintenanceProtocolCUM}   
\end{figure*}



\noindent{\bf The ${\sf write}()$ operation}. When the \op{write} operation is invoked, the writer increments $csn \leftarrow csn +_m 1$, sends {\sc write}$(\ang{v}{csn})$ to all servers and finally returns after $\delta$ time.
For each server $s_i$, two cases may occurs, $s_i$ delivers {\sc write}$(\ang{v}{csn})$ message when it is not affected by a Byzantine agent or when it is affected by a Byzantine agent. In the first case $s_i$ stores $v$ in $W_i$ and forwards it to every server sending the {\sc echo}$(i, \ang{v}{csn}, pending\_read_i)$ message. Such value is further echoed at the beginning of each next \op{maintenance} operation as long as $\ang{v}{csn}$ is in $W_i$ or $V_i$, this is true for $\#echo_{CUM}$ correct servers. When $\ang{v}{csn}$ occurs $\#echo_{CUM}$ times in $echo\_vals_i$ then $s_i$ tries to update $V_{safe_i}$ set by invoking ${\sf insert}$ on the value returned by the ${\sf select\_pairs} (echo\_vals_i)$ function.\\ 

\noindent{\bf The ${\sf read}()$ operation}. At client side, when the \op{read} operation is invoked at client $c_i$, it empties the $reply_i$ set and sends to all servers the {\sc read}$(i)$ message. Then $c_i$ waits $3\delta$ time, while the $reply_i$ set is populated with servers replies, and from such set it picks the newest value occurring $\#echo_{CUM}$ times invoking ${\sf select\_value}(reply_i)$ and returns it. Notice that before returning $c_i$ sends to every server the read termination notification, {\sc read\_ack}$(i)$ message. At server side when $s_j$ delivers the {\sc read}$(i)$ message, client $c_i$ identifier is stored in the $pending\_read_j$ set. Such set is part of the content of {\sc echo} message in every \op{maintenance} operation, which populates the $echo\_read_j$ set, so that cured servers can be aware of the reading clients. Afterwards, $s_j$ invokes ${\sf conCut}(V_j, V_{safe_i},W_i)$ function to prepare the reply message for $c_i$. The result of such function is sent back to $c_i$ in the {\sc reply} message. 
Finally a {\sc reply} message containing just one value is sent when a new value is added in $W_i$ and there are clients in the $pending\_read_j \cup echo\_read_j$ set. When the {\sc read\_ack}$(i)$ message is delivered from $c_i$ then its identifier is removed from the $pending\_read_j$ and $ echo\_read_j$ sets.

\begin{figure*}[!]
\centering
\fbox{
\begin{minipage}{0.4\textwidth}
\scriptsize
\resetline
\begin{tabbing}
aaaA\=aA\=aA\=aaaA\kill

{\bf operation} ${\sf write}(v)$: ~~~~~~~~~~~~~~~~~~~~~~~~~~~~~~~~~ \\

~\\

\line{W-CCUM-01} \> $csn \leftarrow (csn+_m 1)$;\\

\line{W-CCUM-02} \> ${\sf broadcast}$ {\sc write}$(v, csn)$;\\

\line{W-CCUM-03} \> {\bf wait} $(\delta)$;\\

\line{W-CCUM-04} \> {\bf return} ${\sf write\_confirmation}$;

\end{tabbing}
\normalsize
\end{minipage}%
}
\label{sfig:writeCCUM}
\fbox{
\begin{minipage}{0.4\textwidth}
 
\scriptsize
 
\begin{tabbing}
aaaA\=aA\=aA\=aaA\kill

{\bf when} \=  {\sc write}$(v, csn)$ \= is  ${\sf received}$: \\ 


\line{W-SCUM-02} \> $W_i \leftarrow W_i \cup \ang{\ang{v}{csn}}{setTimer(2\delta)}$;\\

\line{W-SCUM-03} \> ${\sf broadcast}$ {\sc echo}$(i, \ang{v}{csn}, pending\_read_i)$;\\

\line{W-SCUM-04} \>  {\bf for} \= {\bf each} $j \in (pending\_read_i \cup echo\_read_i)$ {\bf do}\\

\line{W-SCUM-05} \> \>  ${\sf send}$ {\sc reply} $(i, \{\langle v, csn \rangle\} )$;\\

\line{W-SCUM-06} \> {\bf endFor}

\end{tabbing}
\normalsize
\end{minipage}
}
\label{sfig:writeSCUM}

\caption{$\mathcal{A}_{W}$ algorithms, server side and client side respectively, implementing the ${\sf write}(v)$ operation in the $(\Delta S, CUM)$ model with bounded timestamp.}
\label{fig:writeProtocolCUM}   
\end{figure*}



\begin{figure}[!]
\centering
\fbox{
\begin{minipage}{0.4\textwidth}
\scriptsize
\resetline
\begin{tabbing}
aaaA\=aA\=aA\=aaaA\kill

{\bf operation} ${\sf read}()$: ~~ \\


\line{R-CCUM-01} \> $reply_i \leftarrow \emptyset$;\\

\line{R-CCUM-02} \> ${\sf broadcast}$ {\sc read}$(i)$;\\

\line{R-CCUM-03} \> {\bf wait }$(3\delta)$;\\

\line{R-CCUM-04} \> $\langle v, sn \rangle \leftarrow {\sf select\_value}(reply_i)$;\\

\line{R-CCUM-05} \> ${\sf broadcast}$ {\sc read\_ack}$(i)$;\\

\line{R-CCUM-06} \> {\bf return} $v$;\\

~-------------------------------------------------------------\\

{\bf when} \=  {\sc reply} $(j, V_{set})$ \= is  ${\sf received}$: \\ 


\line{R-CCUM-07} \> {\bf for each}\=  $(\langle v, sn \rangle \in V_{set})$ {\bf do}\\

\line{R-CCUM-08} \>\> $reply_i \leftarrow reply_i \cup \{\langle v, sn \rangle_j \}$;\\

\line{R-CCUM-09} \> {\bf endFor}

%
%

\end{tabbing}
\normalsize
\end{minipage}%
\label{sfig:readCCUM}
}
\fbox{
\begin{minipage}{0.4\textwidth}
 
\scriptsize
 
\begin{tabbing}
aaaA\=aA\=aA\=aaA\kill

{\bf when} \=  {\sc read} $(j)$ \= is  ${\sf received}$: \\ 

\line{R-SCUM-01} \> $pending\_read_i \leftarrow pending\_read_i \cup \{j\}$;\\

\line{R-SCUM-02} \> ${\sf send}$ {\sc reply} $(i, {\sf conCut}(V_i,V_{safe_i},W_i))$;\\

\line{R-SCUM-03} \> ${\sf broadcast}$ {\sc read\_fw}$(j)$;\\

~-----------------------------------------------------------------------\\

{\bf when} \=  {\sc read\_fw} $(j)$ \= is  ${\sf received}$: \\ 

\line{R-SCUM-04} \> $pending\_read_i \leftarrow pending\_read_i \cup \{j\}$;\\

~-----------------------------------------------------------------------\\

{\bf when} \=  {\sc read\_ack} $(j)$ \= is  ${\sf received}$: \\ 

\line{R-SCUM-05}\>$pending\_read_i \leftarrow pending\_read_i \setminus \{j\}$;\\

\line{R-SCUM-06} \>$echo\_read_i \leftarrow echo\_read_i \setminus \{j\}$;\\

\end{tabbing}

\normalsize
\end{minipage}
\label{sfig:readSCUM}
}
\caption{$\mathcal{A}_{R}$ algorithms, client side and server side respectively, implementing the ${\sf read}()$ operation in the $(\Delta S, CUM)$ model with bounded timestamp.}
\label{fig:readProtocolCUM}   
\end{figure}

\subsection{Correctness proofs}\label{sec:proof}
\setcounter{theorem}{0}
In the following we prove that the protocol defined in Section \ref{sec:algorithm} is correct. 

\begin{definition}[Faulty servers in the interval $I$]
Let us define as $\tilde{B}[t,t+T]$ the set of servers that are affected by a Byzantine agent for at least one time unit in the time interval $[t,t+T]$. More formally  $\tilde{B}[t,t+T]= \bigcup_{\tau \in [t,t+T]} B(\tau)$.
\end{definition}

\begin{definition}[$Max\tilde{B}(t,t+T)$]
	Let $[t,t+T]$ be a time interval. The cardinality of $\tilde{B}(t,t+T)$ is maximum if for any $t'$, $t'>0$, is it true that $|\tilde{B}(t,t+T)|\geq |\tilde{B}(t',t'+T)|$. Let $Max\tilde{B}(t,t+T)$ be such cardinality.
\end{definition}

\begin{lemma}\label{l:MaxB}
	If $\Delta > 0$ and $T \geq \delta$ then $Max\tilde{B}(t,t+T)=(\lceil \frac{T}{\Delta} \rceil +1)f$. 
\end{lemma}

\begin{proofL}	
	For simplicity let us consider a single agent $ma_1$, then we extend the reasoning to all the $f$ agents. In the $[t,t+T]$ time interval, with $T \geq \delta$, $ma_1$ can affect a different server each $\Delta$ time. It follows that the number of times it may \vir{jump} from a server to another is $ \frac{T}{\Delta}$. Thus the affected servers are at most $\lceil \frac{T}{\Delta} \rceil$ plus the server on which $ma_1$ is at $t$. Finally, extending the reasoning to $f$ agents, $Max\tilde{B}(t,t+T)=(\lceil \frac{T}{\Delta} \rceil +1)f$, concluding the proof.	   
	\renewcommand{\toto}{l:MaxB}
\end{proofL}

In the following we first characterize the correct system behavior, i.e., when the protocol is correctly executed after $\tau_{stab}$ (the end of the transient failure and system stabilization). In doing this we assume that it is always possible to establish the correct order among the values that are collected by clients and servers. After we prove that it is always possible to establish an order among those values and finally, we prove that the protocol is self-stabilizing after a finite number of \op{write} operations.\\
Concerning the protocol correctness, the termination property is guaranteed by the way the code is designed, after a fixed period of time all operations terminate. The validity property is proved with the following steps:
\begin{itemize}
\item[1.] \op{maintenance} operation works (i.e., at the end of the operation $n-f$ servers store valid values). 
In particular, for a given value $v$ stored by $\#echo_{CUM}$ correct servers at the beginning of the \op{maintenance} operation, there are $n-f$ servers that store $v$ after $\delta$ time since the beginning of the operation;
\item[2.] given a \op{write} operation that writes $v$ at time $t$ and terminates at time $t+\delta$, there is a time $t'< t+3\delta$ after which $\#reply_{CUM}$ correct servers store $v$;
\item[3.] at the next \op{maintenance} operation after $t'$ there are $\#reply_{CUM}-f=\#echo_{CUM}$ correct servers that store $v$, for step (1) this value is maintained in the register;
\item[4.] the validity property follows considering that the \op{read} operation is long enough to include the $t'$ of the last written value in such a way that servers have enough time to reply and after $t'$ this value is maintained in the register, step (3), as long as there are no others \op{write} operations. To such purpose we show that $V_i$ is big enough to do not be full filled with new values before that the last written value is returned.
\end{itemize}

\subsection*{Correctness proofs considering $t>t_{stab}$.}
In the following we prove the correctness of the protocol when there are no transient failures, the system is stable and thus timestamp are not bounded, thus it is always possible to uniquely order all values.

\begin{lemma}\label{lem:wTermCUM}
	If a client $c_i$ invokes a ${\sf write}(v)$ operation at time $t$ then this operation terminates at time $t+\delta$.
\end{lemma}

\begin{proofL}
	The claim simply follows by considering that a ${\sf write\_confirmation}$ event is returned to the writer client $c_i$ after $\delta$ time, independently of the servers behavior (see lines \ref{W-CCUM-03}-\ref{W-CCUM-04}, Figure \ref{fig:writeProtocolCUM}).
	\renewcommand{\toto}{lem:wTermCUM}
\end{proofL}

\begin{lemma}\label{lem:rTermCUM}
	If a client $c_i$ invokes a ${\sf read}()$ operation at time $t$ then this operation terminates at time $t+3\delta$.
\end{lemma}

\begin{proofL}
	The claim simply follows by considering that a ${\sf read}()$ returns a value to the client after $3\delta$ time, independently of the behavior of the servers (see lines \ref{R-CCUM-03}-\ref{R-CCUM-06}, Figure \ref{fig:readProtocolCUM}).
	\renewcommand{\toto}{lem:rTermCUM}
\end{proofL}


\begin{theorem}\label{th:termination}
	Any operation invoked on the register eventually terminates.
\end{theorem}

\begin{proofT}
	The proof simply follows from Lemma \ref{lem:wTermCUM} and Lemma \ref{lem:rTermCUM}.
	\renewcommand{\toto}{th:termination}
\end{proofT}


\begin{lemma}[Step 1.]\label{lem:maintenanceNoWriteCUM}
	Let $v$ be a value stored at $\#echo_{CUM}$ correct servers $s_j \in Co(T_i)$, $v \in V_j \forall s_j \in Co(T_i)$. Then $\forall s_c \in Cu(T_i)$ at $T_i+\delta$ (i.e., at the end of the \op{maintenance}) $v$ is returned by the function  ${\sf select\_pairs}(echo\_vals_i)$.  
\end{lemma}

\begin{proofL} 
	By hypothesis at $T_i$ there are $\#echo_{CUM}$ correct servers $s_j$ storing the same $v$ and running the code in Figure \ref{fig:stateMaintenanceProtocolCUM}. In particular each server broadcasts a {\sc echo}$()$ message with attached the content of $V_j$ which contains $v$ (line \ref{M-CUM-06}). 
	Messages sent by $\#echo_{CUM}$ correct servers are delivered by $s_c$ and stored in $echo\_vals_c$. The communication channels are synchronous, thus  by time $Ti+\delta$ function ${\sf select\_pairs}(echo\_vals_c)$ returns $v$. 
\renewcommand{\toto}{lem:maintenanceNoWriteCUM}
\end{proofL}

\begin{lemma}\label{lem:curedDoNotHarm}
	Let $s_i$ be a correct server running the \op{maintenance} operation at time $T_i$, then if $v$ is returned by the function ${\sf select\_pairs}(echo\_vals_i)$ there exist a \op{write} operation that wrote such value. 
\end{lemma}

\begin{proofL}
Let us suppose that ${\sf select\_pairs}(echo\_vals_i)$ returns $v'$ and  there no exist a {\sf write}$(v')$. This means that $s_i$ collects in $echo\_vals_i$ at least $\#echo_{CUM}$ occurrences of $v'$ coming from cured and Byzantine servers. Let us consider a cured server $s_c$ running the \op{maintenance} operation at time $T_c$. At the beginning of the \op{maintenance} operation $s_c$ broadcasts values contained in $V_i$ and $W_i$ (Figure \ref{fig:stateMaintenanceProtocolCUM}, line \ref{M-CUM-06}). $V_i$ is reset at each operation with the content of $V_{safe_i}$ which is reset at each \op{maintenance} operation (line \ref{M-CUM-01}). It follows that $s_c$ broadcasts non valid values contained in $V_i$ only during the \op{maintenance} operation run at $T_c$. Contrarily, values in $W_i$, depending on $k$, are broadcast only at $T_c$ or also at $T_{c+1}$. Let us consider two cases: $k=2$ and $k=3$.\\
\noindent{\bf case $k=2$:} In this case since $\Delta=2\delta$ and the maximum value of the timer associated to a value is $2\delta$, thus each cured server $s_c$ broadcasts a non valid value contained in $W_i$ only during the first \op{maintenance} operation. So, during each \op{maintenance} operation there are $f$ Byzantine servers and $f$ cured servers, those are not enough to send $\#echo_{CUM}=2f+1$ occurrences of $v'$. For Lemma \ref{lem:maintenanceNoWriteCUM} this is the necessary condition to return $v'$ invoking ${\sf select\_pairs}(echo\_vals_i)$, leading to a contradiction.\\
\noindent{\bf case $k=3$:} $\Delta =\delta$ and the maximum value of the timer associated to a value is $2\delta$, thus each cured server $s_c$ broadcasts a non valid value contained in $W_i$ during the two subsequent \op{maintenance} operations. Summing up, during each \op{maintenance} operation at time $T_i$  there are $f$ Byzantine servers, $f$ cured servers and $f$ servers that were cured during the previous operation. Those servers are not enough to send $\#echo_{CUM}=3f+1$ occurrences of $v'$, for Lemma \ref{lem:maintenanceNoWriteCUM} this is the necessary condition to return $v'$ invoking ${\sf select\_pairs}(echo\_vals_i)$, leading to a contradiction and concluding the proof.
\renewcommand{\toto}{lem:curedDoNotHarm}
\end{proofL}

From the reasoning used in this Lemma, the following Corollary follow.

\begin{corollary}\label{cor:Wpermanenza}
	Let $s_i$ be a non faulty process and $v$ a value in $W_i$. Such value is in $W_i$ during at most $k-1$ sequential \op{maintenance} operations. 
\end{corollary} 

Finally, considering that servers reply during a \op{read} operation with values in $W_i$, $V_i$ and $V_{safe_i}$. $V_{safe_i}$ is safe by definition, $V_i$ is reset after the first \op{maintenance} operation then it follows that servers can be in a cured state for $2\delta$ time, the time that never written values can be present in $W_i$.

\begin{corollary}\label{c:gammaCUM}
Protocol $\mathcal{P}$ implements a \op{maintenance} operation that implies $\gamma \leq 2\delta$.
\end{corollary}

\begin{lemma}\label{lem:curedDoNotHarmWrite}
Let $T_c$ be the time at which $s_c$ becomes cured. Each cured server $s_c$ can reply back with incorrect message to a \msg{read} message during a period of $2\delta$ time.
\end{lemma}

\begin{proofL}
The proof directly follows considering that the content of a \msg{reply} message comes from the $V_c,V_{safe_c}$ and $W_i$ sets. The first one is filled with the content of $V_{safe_c}$ at the beginning of each \op{maintenance} operation and after $\delta$ time is reset (cf. Figure \ref{fig:stateMaintenanceProtocolCUM}, lines \ref{M-CUM-06a}-\ref{M-CUM-06b}). The second one is emptied at the beginning of each \op{maintenance} operation and the third one keeps its value during $k$ \op{maintenance} operations (cf. Corollary \ref{cor:Wpermanenza}). Thus by time $T_c+2\delta$ $s_c$ cleans all the values that could come from a mobile agent. 
\renewcommand{\toto}{lem:curedDoNotHarmWrite}
\end{proofL}

\begin{lemma}[Step 2.]\label{lem:writeCompletionCUM}
	Let $op_W$ be a ${\sf write}(v)$ operation invoked by a client $c_k$ at time $t_B(op_W)=t$ then at time $t+3\delta$ there are at least $n-2f\geq\#reply_{CUM}$ correct servers such that $v\in V_{safe_i}$ and is returned by the function ${\sf concCut}()$.
\end{lemma}

\begin{proofL}
Due to the communication channel synchrony, the {\sc write} messages from $c_k$ are delivered by servers within the time interval $[t,t+\delta]$; any non faulty server $s_j$ executes the correct algorithm code. When $s_j$ delivers a {\sc write} message it stores the value in $W_j$ and sets the associated timer to $2\delta$ (line \ref{W-SCUM-02}, Figure \ref{fig:writeProtocolCUM}). \\
For Lemma \ref{l:MaxB} in the $[t,t+\delta]$ time interval there are maximum $2f$ Byzantine servers, thus at $t+\delta$ $v$ is stored in $W_j$ at $n-2f\geq\#echo_{CUM}$ correct servers $s_j$. All those servers broadcast $v$ by time $t+\delta$, so by time $t+2\delta$ there are $\#echo_{CUM}$ occurrences of $v$ in $echo\_vals_i$, each server $s_i$ stores $v$ in $V_{safe_i}$. If a Byzantine agent movement happens before $t+2\delta$, i.e., $T_i \in [t+\delta,t+2\delta]$ then at time $T_i$, due to Byzantine agents movements, there are $n-3f\geq\#echo_{CUM}$ correct servers that run the \op{maintenance} operation and broadcast $v$. Thus at time $t+3\delta$, for Lemma \ref{lem:maintenanceNoWriteCUM}, all correct servers are storing $v \in V_{safe_i}$ and by construction $v$ is returned by the function ${\sf conCut}()$. We conclude the proof by considering that there are at least $n-2f\geq \#reply_{CUM}$.
\renewcommand{\toto}{lem:writeCompletionCUM}
\end{proofL}

For simplicity, from now on, given a \op{write} operation $op_W$ we call $t_B(op_W)+3\delta=t_{wP}$ the {\bf persistence time} of $op_W$, the time at which there are at least $\#reply_{CUM}$ servers $s_i$ storing the value written by $op_W$ in $V_{safe_i}$.

\begin{lemma}[Step 3$/$1.]\label{lem:writeStableCUM}
Let $op_W$ be a \op{write} operation and let $v$ be the written value. If there are no other \op{write} operations, the value written by $op_W$ is stored by all correct servers forever (i.e., $v$ is returned invoking the ${\sf conCut}()$ function).
\end{lemma}

\begin{proofL}
From Lemma \ref{lem:writeCompletionCUM} at time $t_{wP}$ there are at least $n-2f\geq\#reply_{CUM}$ correct servers $s_j$ that have $v$ in $V_{safe_i}$. At the next Byzantine agents movement there are $n-2f-f\geq\#echo_{CUM}$ correct server storing $v$ in $V_{safe_i}$, which is moved to $V_i$ and broadcast during the \op{maintenance} operation. For Lemma \ref{lem:maintenanceNoWriteCUM}, after $\delta$ time, all non Byzantine servers are storing $v$ in $V_{safe_i}$. At the next Byzantine agents movement there are $f$ less correct servers that store $v$ in $V_{safe_i}$, but those servers are still more than $\#echo_{CUM}$. It follows that cyclically before each agent movement there are $f$ servers more that store $v$ thanks to the \op{maintenance} and $f$ servers that lose $v$ because faulty, but this set of non faulty servers is enough to successfully run the \op{maintenance} operation (cf. Lemma \ref{lem:maintenanceNoWriteCUM})). By hypothesis there are no more \op{write} operations, then $v$ is never overwritten and all correct servers store $v$ forever.\\  
\renewcommand{\toto}{lem:writeStableCUM}
\end{proofL}

\begin{lemma}[Step 3$/$2.]\label{lem:permanenzaWriteCUM}
Let $op_{W_{0}}, op_{W_1}, \dots, op_{W_{k-1}},$ $ op_{W_k}, op_{W_{k+1}}, \dots$ be the sequence of \op{write} operation issued on the regular register. Let us consider a generic $op_{W_k}$, let $v$ be the written value by such operation and let $t_{wP}$ be its persistence time. Then $v$ is in the register (there are $\#reply_{CUM}$ correct servers storing it) up to time at least $t_B{W_{k+3}}$. 
\end{lemma}

\begin{proofL}
The proof simply follows considering that:
\begin{itemize}
\item for Lemma \ref{lem:writeStableCUM} if there are no more \op{write} operation then $v$, after $t_{wP}$, is in the register forever;
\item any new written value eventually is stored in ordered set  $V_{safe}$, whose dimension is $3$;
\item \op{write} operation occur sequentially.
\end{itemize}
It follows that after $3$ \op{write} operations, $op_{W_{k+1}},op_{W_{k+2}},op_{W_{k+3}}$, $v$ is no more stored in the regular register. 
\renewcommand{\toto}{lem:permanenzaWriteCUM}
\end{proofL}

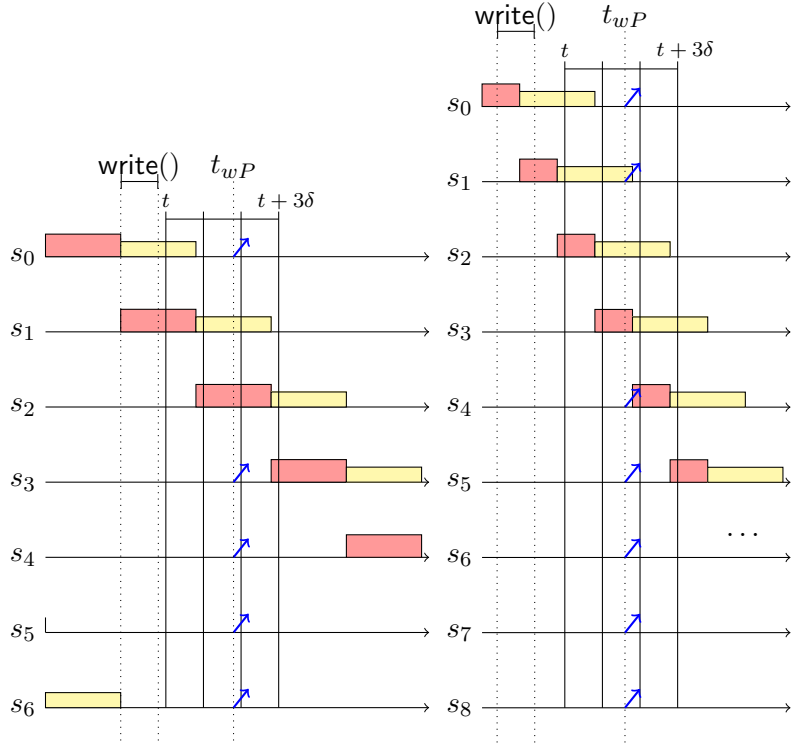
\begin{figure}
\centering
\begin{tikzpicture}[y=-1cm]
\def \lenght {5.1}
\def \n {6} 

\def \deltaGrande {1}
\def \deltaPiccolo {.5}
\def \gammaCuring {1}

\processes
\faults{0}{0}{4}{yellow}{1}
\intervallo{1.6}{3}{t}
\reply{2.5}{0}{blue}
\reply{2.5}{3}{blue}
\reply{2.5}{4}{blue}
\reply{2.5}{5}{blue}
\reply{2.5}{6}{blue}
\scrittura{1}
\lineaVerticale{2.5}{-1}{$t_{wP}$}{dotted}
\end{tikzpicture} %
\begin{tikzpicture}[y=-1cm]
\def \lenght {4.1}
\def \n {8} 

\def \deltaGrande {.5}
\def \deltaPiccolo {.5}
\def \gammaCuring {1}

\processes
\faults{0}{0}{5}{yellow}{0}
\curatoParziale{6*\deltaGrande}{5}{\lenght-6*\deltaGrande-.1}{yellow}
\node[] () at (7*\deltaGrande, 5.7) {$\dots$};
\intervallo{1.1}{3}{t}	
\reply{1.9}{0}{blue}
\reply{1.9}{1}{blue}
\reply{1.9}{4}{blue}
\reply{1.9}{5}{blue}
\reply{1.9}{6}{blue}
\reply{1.9}{7}{blue}
\reply{1.9}{8}{blue}
\scrittura{.2}
\lineaVerticale{1.9}{-1}{$t_{wP}$}{dotted}
\end{tikzpicture}
\caption{In the first scenario $\Delta=2\delta$ and in the second one is $\Delta=\delta$. In red the period during which servers are faulty and in yellow the period during which servers are in a cured state. Blue arrows are the correct replies sent back by correct servers.}\label{fig:scenarioCUMNormale}
\end{figure}

Before to prove the validity property, let us consider how many Byzantine and cured servers can be present during a \op{read} operation that last $3\delta$. For simplicity, to do that we refer to the scenarios depicted in Figure \ref{fig:scenarioCUMNormale}. If $k=3$ there can be up to $4f$ (cf. Lemma \ref{l:MaxB}) Byzantine servers and $2f$ cured servers. If $k=2$ there can be up to $3f$ Byzantine servers (cf. Lemma \ref{l:MaxB}) and $f$ cured servers. In Figure \ref{fig:scenarioCUMNormale} we depicted the extreme case in which there is a \op{read} operation just after the last \op{write} operation. The line marked as $t_{wP}$ represents the time at which for sure correct servers are storing and thus replies with the last written value (cf. Lemma \ref{lem:writeCompletionCUM}). Notice that when $\delta=\Delta$ $s_4$ has just the time to correctly reply to the client before being affected. Notice that if $t_{wP}$ was concurrent with Byzantine agents movements, then during $[t,t+\delta]$ $s_4$ was still able to reply with the last written value because still present in $W_i$, i.e., the reply message happens before the $2\delta$ timer expiration. In any case there are $\#reply_{CUM}$ correct servers that reply with the last written value and the number of those replies is greater than the number of replies coming from cured and Byzantine servers. From those observations the next Corollary follows.

\begin{corollary}\label{c:numeroReplyCorrette}
Let $c_i$ be a client that invokes a \op{read} operation that lasts $3\delta$ time. During such time, the number of replies coming from correct servers is strictly greater than the number of replies coming from Byzantine and cured servers.
\end{corollary}

\begin{theorem}[Step 4.]\label{th:validityCUM}
	Any ${\sf read}()$ operation returns the last value written before its invocation, or a value written by a ${\sf write}()$ operation concurrent with it.
\end{theorem}

\begin{proofT}
Let us consider a \op{read} operation $op_R$. We are interested in the time interval $[t_B(op_R),$ $ t_B(op_R)+\delta]$. The operation lasts $3\delta$, thus reply messages sent  by correct servers within  $t_B(op_R)+2\delta$ are delivered by the reading client. During $[t,t+2\delta]$ time interval there are at least $\#reply_{CUM}$ correct servers that have the time to deliver the read request and reply (cf. Corollary \ref{c:numeroReplyCorrette}). We have to prove that what those correct servers reply with is a valid value. 
There are two cases, $op_R$ is concurrent with some \op{write} operations or not. \\
{\bf - $op_R$ is not concurrent with any \op{write} operation}. Let $op_W$ be the last \op{write} operation such that $t_E(op_W)\leq t_B(op_R)$ and let $v$ be the last written value. For Lemma \ref{lem:writeStableCUM} after the write persistence time $t_{wP}$ there are at least $\#reply_{CUM}$ correct servers storing $v$ (i.e., $v \in {\sf conCut}(V_i, V_{safe_i},W_i)$. Since $t_B(op_R)+2\delta \geq t_Cw$, then there are $\#reply_{CUM}$ correct servers replying with $v$. So the last written value is returned.\\
{\bf - $op_R$ is concurrent with some \op{write} operation}. Let us consider the time interval $[t_B(op_R),$ $ t_B(op_R)+2\delta]$. In such time there can be at most three sequential \op{write} operations $op_{W_1},op_{W_2},op_{W_3}$. Let $op_{W_0}$ be the last write operation before $op_R$. In the extreme case in which those operations happen one after the other we have the following situation. $t_E(op_{W_0}<t_B(op_R))$ and the write persistence time of $op_{W_0}$, $t_{wP_0}<t_B(op_{W_0})+3\delta<t_B(op_R)+2\delta <t_B(op_{W_3})$. Basically, the value written by $op_{W_0}$ is overwritten in $V_i$ by the value written $op_{W_3}$, but not before $t_B(op_R)+2\delta$, thus all correct servers have the time to reply with the last written value. 
Notice that the concurrently written values may be returned if the \msg{write} and \msg{reply} messages are fast enough to be delivered before the end of the \op{read} operation.
To conclude, for Lemma \ref{lem:curedDoNotHarmWrite} Byzantine and cured servers can no force correct servers to store and thus to reply with a never written value. Only cured and Byzantine servers can reply with non valid values. As we stated, if $k=2$ there are up to $4f$ non correct servers. If $k=3$ there are $6f$ non correct servers. In both cases the threshold $\#reply_{CUM}$ is higher than the occurrences of non valid values that a reader can deliver. Mobile agents can not force the reader to read another or older value and even if an older values has $\#reply_{CUM}$ occurrences the one with the highest sequence number is chosen.  
\renewcommand{\toto}{th:validityCUM}
\end{proofT}

From the reasoning used to prove Theorem \ref{th:validityCUM} the next Corollary follows.

\begin{corollary}\label{c:lastWrittenValue}
When a client $c_i$ invokes a \op{read} operation the last written value occurs in $reply_i$ at least $\#reply_{CUM}$ times.
\end{corollary}

\begin{theorem}\label{t:CUM}
	Let $n$ be the number of servers emulating the register and let $f$ be the number of Byzantine agents in the $(\Delta S, CUM)$ round-free Mobile Byzantine Failure model with no transient failures.
	Let $\delta$ be the upper bound on the communication latencies in the synchronous system.
	If (i) $n\geq 6f+1$ for $\Delta = 2\delta$ and (ii) $n \ge 8f+ 1$ for $\Delta = \delta$, then $\mathcal{P}_{reg}$ implements a SWMR Regular Register in the $(\Delta S, CUM)$ round-free Mobile Byzantine Failure model.
\end{theorem}

\begin{proofT}
	The proof simply follows from Theorem \ref{th:termination} and Theorem \ref{th:validityCUM}.
	\renewcommand{\toto}{t:CUM}
\end{proofT}

\subsection*{Self-Stabilization correctness proofs}

What is left to prove are the necessary conditions for the system to self-stabilize after $\tau_{no\_tr}$. We first prove that with timestamps in $\mathcal{Z}_{13}$ it is always possible to uniquely order the values that clients and servers manage at the same time. Then, we prove that when the system is not stable, given the fact the timestamps are bounded, after a finite number of \op{write} operations the system becomes stable.  

\begin{lemma}\label{lem:orderDuringWrite}
During each \op{write} operation $op_W$ such that $t_B(op_W)>\tau_{stab}$, each non faulty server $s_i$ has at most $6$ values returned by the function ${\sf select\_pairs} (echo\_vals_i)$ during the same \op{maintenance} operation and it is always possible to uniquely order them. 
\end{lemma}

\begin{figure}
	\begin{tikzpicture}[y=-1cm]
		\def \lenght {3}
		\def \n {2} 
		
		\def \deltaGrande {.5}
		\def \deltaPiccolo {.7}
		\def \gammaCuring {1}

		\processes
		\node[] () at (0,3) {$\dots$};
		\foreach \t in {0,...,2}{
			\foreach \x in {0,.7, 1.4,...,2.1}{
				\draw [|-|] (\x,\t)--(\x+.7, \t);
			}
		}

		\scritturaValore{.5}{3}
		\scritturaValore{1.2}{4}
		\scritturaValore{1.9}{5}
		
		\node[] () at (2.1,-.2) {\scriptsize \color{red} (a)};
	\end{tikzpicture}
	\hspace{.2cm}
  \begin{tikzpicture}[y=-1cm, rotate=90, scale=.5]
  	\def \angolo {27.6}
    \draw[black, thick] (0,0) circle (2);    
    \foreach \a in {0, \angolo,...,360}{
    	\draw[thick,black] (0, 0) -- (\a:2.3);
    }
    \foreach \b in {0,...,12}{
   		\draw (\angolo*\b: 3) node {\scriptsize \b};
    
    }
    \draw[fill, red] (0: 2) circle (.1cm);
    \draw[fill, red] (\angolo*1: 2) circle (.1cm); 
    \draw[fill, red] (\angolo*2: 2) circle (.1cm);
    \draw[fill, red] (\angolo*5: 2) circle (.1cm);  
    \draw[fill=black] (0,0) circle(0.7mm);
    \draw [-latex] (-1,.2) arc [start angle=200, end angle=-130, x radius=1cm, y radius=1cm];
    \draw [-latex, red, dotted] (-1.5,-2.2)  arc [start angle=150, end angle=10, x radius=1.5cm, y radius=.7cm];
    \draw [-latex] (1.5,2.2) arc [start angle=150, end angle=0, x radius=-1.5cm, y radius=-.7cm];
  \end{tikzpicture}

\caption{Example for $\Delta=\delta$. The small vertical lines are the points where \op{maintenance} operations terminates and begins. Servers are storing $V_i=\{0,1,2\}$, for simplicity we represent values with their timestamp and we consider only correct servers.}\label{fig:writeExample}
\end{figure}
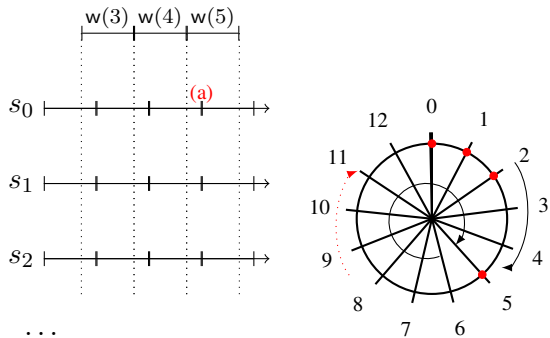

\begin{proofL}
For sake of simplicity let us consider the scenario depicted in Figure \ref{fig:writeExample}, where $op_W(3),$ $op_W(4),op_W(5)$, a sequence of \op{write} operations, occurs (we represent each value with its associated timestamp $\in \mathcal{Z}_{13}$). By hypothesis the system is stable $t_B(op_W)>\tau_{stab}$, thus each correct server $s_i$ has $V_i=\{0,1,2\}$ and those values are broadcast at the beginning of the \op{maintenance} operation along with values in $W_i$ as long as there are not enough occurrences of those values to be stored in $V_{safe_i}$ and then $V_i$. 
Considering that:
\begin{itemize}
\item from Lemma \ref{lem:writeCompletionCUM}, for each \op{write} operation $op_W$, by time $t<t_B(op_W)+3\delta$ the written value is stored in $V_{safe_i}$;
\item a written value $v$ is removed from $W_i$ after $2\delta$ time, so at most by time $t_B(op_W)+3\delta$ $v\notin W_i$;
\item \op{write} operations are sequential and last $\delta$ time (cf. \ref{lem:wTermCUM})
\item $V_{safe_i}$ is a $3$ dimension set.
\end{itemize}
It follows that given a sequence of at least three \op{write} operations $op_W(3),op_W(4),op_W(5)$, before that $op_W(5)$ terminates the value written by $op_W(3)$ is in $V_{safe_i}$ and $0$ has been discharged (cf. Figure \ref{fig:writeExample} the point marked by (a)), i.e. $t_E(op_W(5)>t_B(op_w(3)+3\delta)$. It follows if $op_W(6)$ occurs such that $t_B(op_W(6))=t_E(op(W(5)))$ then $0$ is no more in $V_i$ and during this time $1$ is overwritten by $3$. Generalizing, during each \op{maintenance} operation there are at most the $3$ values in $V_i$ and values belonging to the last $3$ \op{write} operations. \\
Let us now prove the second part of the statement. 
Considering that:
\begin{itemize}
\item timestamps are generated sequentially;
\item during the same \op{maintenance} operation timestamps can span a range of $6$ values.
\end{itemize}
Then for each couple of timestamp $ts_q$ and $ts_p$ returned by ${\sf select\_pairs} (echo\_vals_i)$ during the \op{maintenance} operation, if $ts_q$ has been generated before $ts_p$ then $ts_p-_m ts_q \leq 5$. This means that given $\mathcal{Z}_{13}$ and $ts_q,ts_p$ there is only one way to order them. If $ts_p$ is generated before $ts_q$ then $ts_q-_m ts_p \geq 7$ which is a contradiction with the fact that during the same \op{maintenance} operation timestamps can span a range of $6$ values (cf. the \vir{clock} depicted in Figure \ref{fig:writeExample}).
\renewcommand{\toto}{lem:orderDuringWrite}
\end{proofL}

\begin{lemma}\label{lem:orderDuringRead}
During each \op{read} operation $op_R$ such that $t_B(op_R)>\tau_{stab}$, each client $c_i$ delivers at most $9$ values whose occurrence is $\#reply_{CUM}$ and it is always possible to uniquely order them.
\end{lemma}

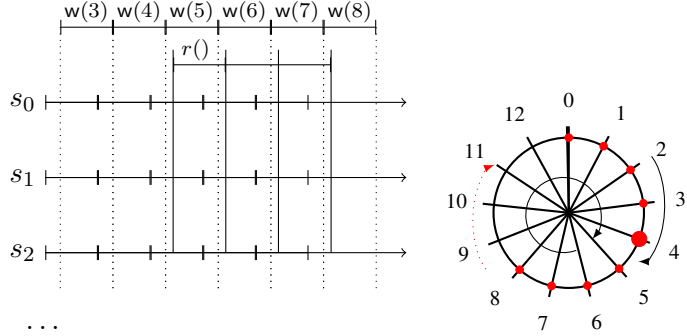
\begin{figure}
  		\begin{tikzpicture}[y=-1cm]
  			\def \lenght {4.8}
  			\def \n {2} 
  			
  			\def \deltaGrande {.5}
  			\def \deltaPiccolo {.7}
  			\def \gammaCuring {1}

  			\processes
  			\node[] () at (0,3) {$\dots$};
  			\foreach \t in {0,...,2}{
  				\foreach \x in {0,.7, 1.4,...,2.8}{
  					\draw [|-|] (\x,\t)--(\x+.7, \t);
  				}
  			}


  			\scritturaValore{.2}{3}
  			\scritturaValore{.9}{4}
  			\scritturaValore{1.6}{5}
  			\scritturaValore{2.3}{6}
  			\scritturaValore{3}{7}
  			\scritturaValore{3.7}{8}
  			\letturaLunga{1.7}{3}
  			\node[] () at (2,-.7) {\scriptsize$r()$};
  			
  		\end{tikzpicture}
  		\hspace{.2cm}
  		\begin{tikzpicture}[y=-1cm, rotate=90, scale=.5]
		 	\def \angolo {27.6}
		    \draw[black, thick] (0,0) circle (2);    
		    \foreach \a in {0, \angolo,...,360}{
		    	\draw[thick,black] (0, 0) -- (\a:2.3);
		    }
		    \foreach \b in {0,...,12}{
		   		\draw (\angolo*\b: 3) node {\scriptsize \b};
		    
		    } 
		    \draw[fill, red] (0: 2) circle (.1cm);
		    \draw[fill, red] (\angolo*1: 2) circle (.1cm); 
		    \draw[fill, red] (\angolo*2: 2) circle (.1cm); 
		    \draw[fill, red] (\angolo*3: 2) circle (.1cm);
		    \draw[fill, red] (\angolo*4: 2) circle (.2cm);
		    \draw[fill, red] (\angolo*5: 2) circle (.1cm);
		    \draw[fill, red] (\angolo*6: 2) circle (.1cm); 
		    \draw[fill, red] (\angolo*7: 2) circle (.1cm);  
		    \draw[fill, red] (\angolo*8: 2) circle (.1cm);    
		    \draw[fill=black] (0,0) circle(0.7mm);
		    \draw [-latex] (-1,.2) arc [start angle=200, end angle=-130, x radius=1cm, y radius=1cm];
		    \draw [-latex, red, dotted] (-1.5,-2.2)  arc [start angle=150, end angle=10, x radius=1.5cm, y radius=.7cm];
		    \draw [-latex] (1.5,2.2) arc [start angle=150, end angle=0, x radius=-1.5cm, y radius=-.7cm];
      	\end{tikzpicture}
 \caption{Example for $\Delta=\delta$. The small vertical lines are the points where \op{maintenance} operations terminates and begins. In both cases servers are storing $V_i=\{0,1,2\}$, for simplicity we represent values with their timestamp and we consider only correct servers.}\label{fig:readExample}
\end{figure}

\begin{proofL}
For simplicity let us consider the scenario depicted in Figure \ref{fig:readExample}, where the \op{read} operation $op_R$ happens after the end of the last \op{write} operations $op_W(3)$ and $op_W(4)$ but before their persistence time $t_{wP}$. Moreover $op_R$ is concurrent with four subsequent \op{write} operations $op_W(4), op_W(5), op_W(6), op_W(7)$. 
During $op_W$ we have the following:
\begin{itemize}
\item $t_{wP}$ of $op_W(3)$ and $op_W(4)$ are after $t_B(op_R)$;
\item for times constraints all the previous \op{write} operations are in $V_i$ at each correct servers;  
\item by hypothesis the system is stable, thus each correct server $s_i$ can have $V_i=\{0,1,2\}$ (no yet overwritten by $op_W(3)$ and $op_W(4)$ values);
\item for Corollary \ref{c:lastWrittenValue} each correct server replies with the last written value $4$;
\item if messages are fast enough each correct server can reply with also $\{5,6,7, 8\}$.   
\end{itemize}
Thus $c_i$ has potentially $9$ values that occur $\#reply_{CUM}$ times.\\
To prove the second part of the statement, consider that for Corollary \ref{c:lastWrittenValue} the last written value is always present. Thus, there can be up to $9$ sequential values $ts_p$ and the last written value $ts_{ls}$ is in the middle, it follows that the distance between each value $ts_p$ and the one in the middle $ts_{ls}$ is such that $|ts_p-_m ts_{lw}|\leq 4$ (cf. the \vir{clock} in Figure \ref{fig:readExample}). If follows that for each triple of value there always exist a value in the middle such that there is one only way to order them. For example, let us consider the set $\{7,0,3\}$ it can be ordered in three ways: $7,0,3$, $3,7,0$ and $0,3,7$. In the first two cases $7-_m 0 >4$ and $7-_m 0 >4$ respectively, thus the order $0,3,7$ is the only possible one.
\renewcommand{\toto}{lem:orderDuringRead}
\end{proofL}

\begin{lemma}\label{lem:writes2stab}
Let $op_{W_1}, \dots, op_{W_{10}}$ be a sequence of 10 \op{write} operations, occurring after $\tau_{no\_tr}$. At time $t>t_E(op_{W_{10}})$ the system is self-stabilized.
\end{lemma}

\begin{proofL}
For Lemma \ref{lem:maintenanceNoWriteCUM} if there are $\#echo_{CUM}$ correct servers storing the same value $v$ then such value is stored by all correct servers after $\delta$ time since the beginning of the \op{maintenance} operation. Thus given the first \op{maintenance} operation after $\tau_{no\_tr}$ correct servers are either storing the same values or empty set. If correct servers are storing nothing, then after the end of the first \op{write} operation the system is stable, since for Corollary \ref{c:lastWrittenValue} such a value is returned by the next \op{read} operation.\\
If $V_{safe_i}$ is not empty then different scenarios may happen.
\begin{itemize}
\item{case a.} If values stored in $V_{safe_i}$ have not an unique order (e.g., $\exists ts_p,ts_q \in \mathcal{Z}_{13}: ts_p=ts_q \vee |ts_p-_m ts_q|>5$), then at the next \op{maintenance} operation, the function ${\sf checkOrderAndTrunc}(V_{safe_i})$ resets such set. Notice that such reset may happen at the beginning of different \op{maintenance} operations, depending on $\Delta$. For sake of simplicity let us consider Figure \ref{fig:writeExample}, $3$ can be stored in $V_{safe_0}$ during the three different \op{maintenance} operations that occur since the beginning of the \op{write} operation and the point marked as (a), the $t_{wP}$. In such time interval  two other \op{write} operations may occur. But after the point marked as (a) all non Byzantine servers reset their $V_{safe_i}$ set. Thus now, after the end of the next \op{write} operation the system is stable, indeed, for Corollary \ref{c:lastWrittenValue} such value is returned by the next \op{read} operation. Thus in such a case after four \op{write} operations the system is stable;
\item{case b.} If values stored in $V_{safe_i}$ can be uniquely ordered, e.g. $V_{safe_i}=\{0,1,2\}$ then three scenarios may happen;
\begin{itemize}
\item the next written value does not have an unique order with respect to each value in $V_{safe_i}$, (e.g., $0,1,2,6,7$) then again the set is reset and case (a.) takes place;
\item  the next written value timestamp is ordered as newest with respect to the values in $V_{safe_i}$ (e.g., $3,4,5$) and then, at the end of the \op{write} operation, the system is stabilized. Indeed for Corollary \ref{c:lastWrittenValue} such value is returned to the next \op{read} operation;
\item the next written value is ordered as older with respect to each value in $V_{safe_i}$ and thus is dropped. This happens up to the \op{write} operation that writes a value equal to a value in $V_{safe_i}$, when this happens $V_{safe_i}$ is reset and after four \op{write} operations the system is stable (cf. case a.). If $V_{safe_i}=\{0,1,2\}$ then in the worst case are needed $6$ \op{write} operations, e.g., $8,9,10,11,12,0$. Then at the next \op{maintenance} operation, when two values associated with $0$ are in $V_{safe_i}$ the function ${\sf checkOrderAndTrunc}(V_{safe_i})$ resets such set. Thus, after the end of the next four \op{write} operations the system is stable (cf. case a.), indeed for Corollary \ref{c:lastWrittenValue} such value is returned by the next \op{read} operation. Thus, after $10$ \op{write} operations the system is stable.
\end{itemize} 
\end{itemize}
Considering all those cases, the worst case scenario happens when $10$ \op{write} operations are required to stabilise the system. The claim trivially follows generalizing the argumentation for general timestamps in $\mathcal{Z}_{13}$.
\renewcommand{\toto}{lem:writes2stab}
\end{proofL}

\begin{theorem}\label{t:ssCUM}
	Let $n$ be the number of servers emulating the register and let $f$ be the number of Byzantine agents in the $(\Delta S, CUM)$ round-free Mobile Byzantine Failure model.
	Let $\delta$ be the upper bound on the communication latencies in the synchronous system.
	If (i) $n\geq 6f+1$ for $\Delta = 2\delta$ and (ii) $n \ge 8f+ 1$ for $\Delta = \delta$, then $\mathcal{P}_{reg}$ implements a Self-Stabilizing SWMR Regular Register in the $(\Delta S, CUM)$ round-free Mobile Byzantine Failure model.
\end{theorem}

\begin{proofT}
	The proof simply follows from Theorem \ref{t:CUM} and Lemma \ref{lem:writes2stab}.
	\renewcommand{\toto}{t:ssCUM}
\end{proofT}

\begin{theorem}\label{th:optimal}
Protocol $\mathcal{P}_{Rreg}$ is optimal with respect to the number of replicas. 
\end{theorem}

\begin{proofT}
The proof follows considering that Theorem \ref{t:ssCUM} proved that $\mathcal{P}_{Rreg}$ implements a Regular Register with the upper bounds provided in Table \ref{tab:summaryCUM}. Those bounds match the lower bounds proved in Theorem 1 in \cite{BDPT17}. In particular such Theorem states that no safe register can be solved if $n_{CUM_{LB}}=[2(Max\tilde{B}(t,t+T_r)+MaxCu(t,t+T_r))-min\tilde{CBC}(t,t+T_r)]f$ where $T_r$ is the upper bound on the {\sf read}$()$ operation duration. Each term can be computed applying Table \ref{t:tabelloneGenerale} \cite{BDPT17} considering $\gamma=2\delta$ (Corollary \ref{c:gammaCUM}). In particular if $\Delta=\delta$ then $n_{CUM_{LB}}=[2(4+2)-4]f=8f$ while if if $\Delta=2\delta$ then $n_{CUM_{LB}}=[2(3+1)-2]f=6f$, concluding the proof.
\renewcommand{\toto}{th:optimal}
\end{proofT}

\begin{table}[]
	\centering
	\caption{Values for a general \op{read} operation that terminates after $3\delta$ time \cite{BDPT17}.}
	\label{t:tabelloneGenerale}
	\begin{tabular}{|c|c|c|c|}
		\hline
		& $Max\tilde{B}(t,t+3\delta)$ & $MaxCu(t)$ & $MaxSil(t,t+3\delta)$  \\
		\hline
		$(\Delta S,CUM)$ & $\ceil{\frac{3\delta}{\Delta}}+1$ & $ \mathcal{R}( \ceil{\frac{3\delta - \epsilon - \ceil{\frac{3\delta}{\Delta}}\Delta +\gamma}{\Delta}})$ & $\ceil{\frac{\gamma+\delta - \epsilon - \ceil{\frac{3\delta}{\Delta}}\Delta }{\Delta}}$  \\
		\hline
	\end{tabular}
	\begin{tabular}{|c|c|}
		\hline
		& $min\tilde{CBC}(t,t+3\delta)$ \\
		\hline
		$(\Delta S,CUM)$ &  $\ceil{\frac{3\delta-\epsilon-\delta}{\Delta}}+\mathcal{R}(\ceil{\frac{3\delta}{\Delta}}-\ceil{\frac{\gamma+\delta}{\Delta}})+(MaxCu(t)-MaxSil(t,t+3\delta))$ \\
		\hline
	\end{tabular}
\end{table}

\section{Concluding remarks}\label{sec:conclusion}

This paper proposed a self-stabilizing regular register emulation in a distributed system where both transient failures and mobile Byzantine failures can occur, and where messages and Byzantine agent movements are decoupled.
The proposed protocol improves existing works on mobile Byzantine failures~\cite{BDPT16c,BDP16,BDPT17} being the first self-stabilizing regular register implementation in a round-free synchronous communication model and to do so it uses bounded timestamps from the $\mathcal{Z}_{13}$ domain to guarantee finite and known stabilization time. 
In particular, the convergence time of our solution is upper bounded by $T_{10write()}$, where $T_{10write()}$ is the time needed to execute ten \emph{complete} \op{write} operations. 
Contrary to the $(\Delta S, CAM)$ model, $(\Delta S, CUM)$ model required to design a longer \op{maintenance} operation (that lasts $2\delta$ time). As a side effect, also the \op{read} operation completion time increased and it has a direct impact on the size of the bounded timestamp domain that characterize the stabilization time. 
However, it is interesting to note that all these improvements have no additional cost with respect to the number of replicas that are necessary to tolerate $f$ mobile Byzantine processes and our solution is optimal with respect to established lower bounds. 

An interesting future research direction is to study upper and lower bounds for \emph{(i)} memory, and \emph{(ii)} convergence time complexity of self-stabilizing register emulations tolerating mobile Byzantine faults. Nevertheless, interesting is the study of optimal \op{maintenance} solutions.

\end{document}